\documentclass{IOS-Book-Article}
\usepackage{times}
\normalfont
\usepackage[T1]{fontenc}
\usepackage{amsthm}
\usepackage{amssymb}
\usepackage{amsfonts}
\theoremstyle{plain}
\newtheorem{thm}{Theorem}[subsection]
\newtheorem*{thm*}{Theorem}
\newtheorem*{G-R}{Gale--Ryser Theorem}

\newtheorem*{claim*}{Claim}

\theoremstyle{definition}
\newtheorem{Ex}{Example}[subsection]
\newtheorem*{Ex*}{Example}

\newtheorem{dfn}[thm]{Definition}
\swapnumbers
\theoremstyle{remark}
\newtheorem{rem}[thm]{Remark}%[subsection]
\newtheorem*{rem*}{Remark}
\begin{document}
\begin{frontmatter}                           % The preamble begins here.

%\pretitle{Pretitle}
\title{Dynamic Symmetry Approach\\ to Entanglement}%
%\thanks{Use IOS-Book-Article.tmpl file as a template.}}
\runningtitle{Entanglement}
%\subtitle{Subtitle}

\author{\fnms{Alexander } \snm{Klyachko}%
\thanks{Correspondence to: A.~Klyachko, Bilkent University, 06800, Bilkent, Ankara, Turkey.
Tel.: +90 312 290 2115; Fax: +90 312 266 4579; E-mail:
klyachko@fen.bilkent.edu.tr}},
%\author[A]{\fnms{Second} \snm{Author}}
%and
%\author[B]{\fnms{Third} \snm{Author}}
\runningauthor{A.~Klyachko}
\address{Bilkent University, Turkey}
%\address[B]{Short Affiliation of Third Author}

% ----------------------------------------------------------------

\begin{abstract}
In this lectures I explain a connection between geometric
invariant theory and entanglement, and give a number of examples
how this approach works.
\end{abstract}

\end{frontmatter}

%\hfill \parbox{8cm}{\it
%Symmetry principles underpin the elegant quantum mechanical
%description in an abstract picture in which statics and dynamics
%are paradoxically conflated in a way that often leave us hovering
%on the boundary between abstract mathematical understanding and
%literal physical misunderstanding.
%}\\[0mm]
%
%\hfill Sir Harold Kroto, Nobel Lecture 1996

%\tableofcontents
%\addtocontents{toc}{\tiny}
%\addcontentsline{toc}{section}{\bf}

%\section{Introduction}

%\subsection{} Everybody knows and nobody understand what is
%entanglement. The very term was coined by E. Sch\"odinger in the
%famous {\it cat paradox} paper \cite{Sch35}, which in turn was
%inspired by no less celebrated Einstein-Pdolsky-Rozen {\it
%gedanken experiment} \cite{EPR}.

\section{Physical background}

\subsection{Classical mechanics}
%{Classical and quantum physics}
Let  me start with classical nonlinear equation
\begin{equation}\label{pendulum}
\frac{d^2\theta}{dt^2}=-\omega^2\sin \theta,\quad
\omega^2=\frac{g}{\ell}
\end{equation}
describing graceful
%wavering
swing of a clock pendulum in a corner of
%18-th century
Victorian drawing room. It has double periodic solution
$$\theta(t)=\theta(t+T)=\theta(t+i\tau),$$
with real period $T$, and imaginary one $i\tau$. Out of this
equation, carefully studied by Legendre, Abel, and Jacobi, stems
the whole theory of elliptic functions.
%This is one of numerous examples of
%fruitful exchange of ideas between physics and mathematics.
%%Currently however their relations are all but frozen.
%\begin{quote}
%{\it ``
%%Mathematics is the part of physics where experiments are cheap. ...
%But in the middle of twentieth century it was attempted
%to divide physics and mathematics. The consequences turn out to be
%catastrophic.''}\hfill V.~Arnold
%\end{quote}
%%\cite{Ar97}.
%%(V. Arnold, On teaching of Mathematics,
%%{\tt http://pauli.uni-muenster.de/~munsteg/arnold.html}).
%
%I use this talk as an opportunity to build more confidence between
%%physicists and mathematicians
%natural scientists in order to avert further development of the
%catastrophe.

%{Classical and quantum physics}
 Physicists are less interested in  mathematical
subtleties, and
%instead of nonlinear equation (\ref{pendulum})
usually shrink
%pendulum
equation (\ref{pendulum}) to linear one
%deal with its {\it linearization }
$$\frac{d^2\theta}{dt^2}=-\omega^2\theta,\quad|\theta|\ll1$$
with simple harmonic solution $\theta=e^{\pm i\omega t}$. This
example outlines  a general feature of classical mechanics, where
linearity appears mainly as a useful approximation.

\subsection{Quantum mechanics}
%\subsection{Physics and geometry}
In striking contrast to this, {\it quantum mechanics} is
 intrinsically linear, and therefore {\it more simple\/} then
classical one,
%\footnote{But it has enormous  difficulties to express itself in harsh macroscopic reality.},
in the same way as analytic geometry of Descartes is simpler then
synthetic geometry of Euclid. As a price for
%linearity
its simplicity quantum mechanics
%encounters with
runs into enormous difficulties to manifest itself in a harsh
macroscopic reality. This is what makes quantum phenomenology so
tricky.

Mathematicians
%meet
encounter a similar problem when try to extract geometrical
gist
%content
%essence
from a mess of coordinate calculations.
%In both cases the main challenge is
%search for
%revealing a hidden meaning.
In both cases the challenge is to
cover formal bonds of mathematical skeleton with flesh of meaning.

As we know from Klein's {\it Erlangen program\/}, the geometrical
meaning
%is rooted
rests upon invariant quantities and properties (w.r. to a relevant
{\it structure group\/} $G$). This thesis effectively reduces
``elementary'' geometry to invariant theory.

As far as physics is concerned, we witnessed its progressive
geometrization in the last decades \cite{Wil, Her88} . To name few
examples: general relativity, gauge theories, from electro-weak
interactions to chromodynamics,
%including electrodynamics and
%chromodynamics, Berry phase etc.
are all
%essentially
geometrical in their ideal essence.
%See more in [{\it Physics and Geometry}].
In this lectures, mostly based on preprint \cite{Kl02}, I
%show that
%extend the list to {\it entanglement}, and study
%set up
explain a connection between geometric invariant theory and
entanglement, and give a number of examples how this approach
works. One can find further  applications in \cite{Kl04,Kl05}.
%investigate its close
%show that it is a
%physical counterpart of {\it geometric invariant theory}. and give
%some examples how this approach works.
%The main objective of this paper is to convince
%the reader that entanglement a physical counterpart geometric
%invariant theory.
%to add to the list the theory of entanglement
%which mathematically equivalent to Geometric
%Invariant Theory
%and to give some examples how this
%geometric
%approach works.

%From Felix Klein's {\it Erlangen Program\/} mathematicians knew
%that geometrical meaning
%%stems from
%%is a loose term for
%comes from invariant quantities and properties.
%%{\it invariant theory\/}.
%A similar observation holds for physics:
%\begin{quote}{\it ``Why should we not go directly to
%invariants? The whole of physics is contained in them. The answer
%is that it would be fine if we could do it. But it is not easy.''}
%\hfill { Bryce~S.~DeWitt}
%\end{quote}

%But it has enormous  difficulties to manifest itself in harsh
%macroscopic reality. This makes quantum phenomenology very tricky.

%\subsection{Quantum margins}
\subsection{Von Neumann picture}
A background of a quantum system $A$ is Hilbert space
$\mathcal{H}_A$, called {\it state space}. Here, by default, the
systems are expected to be {\it finite\/}:
$\dim\mathcal{H}_A<\infty$.
%Here I'll consider only {\it finite systems\/}, for which $\dim\mathcal{H}_A<\infty$.
%Actual
A {\it pure state\/} of the system is given by unit vector
$\psi\in\mathcal{H}_A$, or by projector operator
$|\psi\rangle\langle\psi|$, if the phase %of $\psi$
factor is irrelevant. Classical mixture
$\rho=\sum_ip_i|\psi_i\rangle\langle\psi_i|$ of pure states called
{\it mixed state\/} or {\it density matrix\/}. This is a
nonnegative Hermitian operator
$\rho:\mathcal{H}_A\rightarrow\mathcal{H}_A$ with unit trace
$\mathrm{Tr\,}\rho=1$.
% for {\em pure state\/} or by non negative
%while  Hermitian operator
%$\rho:\mathcal{H}_A\rightarrow\mathcal{H}_A,
%\operatorname{Tr}\rho=1$, called {\em density matrix\/}, for {\em mixed state\/}.
%Pure state $\psi$, up to mostly irrelevant phase factor, may be
%identified with
%%corresponds to
%projection operator $|\psi\rangle\langle\psi|$ onto $\psi$. Hence
%\begin{equation}
%\rho=\mbox{pure}\Longleftrightarrow\operatorname{rk}\rho=1
%\Longleftrightarrow\mathrm{Spec\,}\rho=(1,0,0\cdots,0).
%\end{equation}

{\it An observable\/} of the system $A$ is Hermitian operator
$X_A:\mathcal{H}_A\rightarrow\mathcal{H}_A$. Actual measurement of
$X_A$
%observation
%$X_A$
%over the
%while
upon  the system in state $\rho$ produces a random quantity
$x_A\in\mathrm{Spec\,} X_A$ implicitly determined by expectations
$$\langle f(x_A)\rangle_\rho=\mathrm{Tr\,}(\rho
f(X_A))=\langle\psi|f(X_A)|\psi\rangle$$ for arbitrary function
$f(x)$ on $\mathrm{Spec\,} X_A$ (the second equation holds for
pure state $\psi$). The measurement process puts the system into
an eigenstate $\psi_\lambda$ with the observed eigenvalue
$\lambda\in\mathrm{Spec\,} X_A$.
%Admitting an abuse of language we
Occasionally we use ambiguous  notation   $|\lambda\rangle$ for
the eigenstate with eigenvalue $\lambda$.

\subsection{Superposition principle}
The linearity of quantum mechanics is embedded from the outset in
{\it Schr\"odinger equation} describing time evolution of the
system
\begin{equation}\label{Schrod}
i\hbar\frac{d\psi}{dt}=H_A\psi
\end{equation}
where
%$\psi\in \mathcal{H}$ is {\it state vector} and
$H_A:\mathcal{H}_A\rightarrow\mathcal{H}_A$ is the {\it
Hamiltonian} of the system $A$.
%acting in Hilbert {\it
%state space} $
%\mathcal{H}$ of a quantum system under consideration.
Being linear Schr\"odinger equation
%implies unitary evolution of the state vector has
admits simple solution
\begin{equation}\label{evol}
\psi(t)=U(t)\psi(0),
\end{equation}
where $ U(t)=\exp\left(-\frac i\hbar\int_0^tH_A(t)dt\right)$ is
unitary {\it evolution operator}.
%Therefore $|\psi(t)|$ is a
%constant, usually set to one, but to avoid irrelevant factors we
%mostly dispense with normalization.
%I'll leave state vectors mostly unnormalized.

Solutions of Schr\"odinger equation (\ref{Schrod}) form a linear
space. This observation
%was extended to
is a source of general {\it superposition principle,} which claims
that a normalized linear combination
$$a\psi+b\varphi$$
of realizable physical states $\psi,\varphi$ is again a realizable
physical state (with no recipe how to cook it). This may be the
most important revelation about physical reality after atomic
hypothesis. It is extremely counterintuitive and implies, for
example, that one can set the celebrated Shcr\"odinger cat into
the state
$$\psi=|\mbox{dead}\rangle+|\mbox{alive}\rangle$$
intermediate between death and life. As BBC put it:
%\begin{quote}
{\it ``In quantum mechanics it is not so easy to be or not to
be.''}
%\end{quote}

%Another consequence
%of  the superposition principle
%is identification of state space of {\it composite system,} $AB$
%implies that
%of quantum mechanics
%which implies that state space of composite system $AB$ splits
%into
From the superposition principle it follows that state space of
composite system $AB$ splits into {\it tensor product\/}
$$\mathcal{H}_{AB}=\mathcal{H}_A\otimes\mathcal{H}_B$$
of state spaces of the components, as opposed to {\it direct
product\/} $P_{AB}=P_A\times P_B$ of configuration spaces in
classical mechanics.

\subsection{Consequences of linearity}
%{Classical and quantum realities}
The linearity
%evolution (\ref{evol})
imposes severe restrictions on
possible manipulations with quantum states.
%and how they can manifest themselves in macroscopic reality. The restrictions
%mostly come from {\it classical determinism}, which contends  with
%linearity:
%
%\begin{enumerate}
%\item Macroscopically distinct classical
%states must be orthogonal, since otherwise there would be a
%non-minuscule probability of transition between them.
%\item Hence macroscopically distinct classical states are linear independent.
%\item Therefore a nontrivial linear combination of
%macroscopically distinct classical states is not a classical
%state.
%\end{enumerate}
Here is a couple of examples.
\subsubsection{No-cloning Theorem}
%\subsubsection{}
Let's start with notorious claim
%by W.K.~Wooters \& W.H.~Zurek \cite{WZ82} and D.~Dieks \cite{Di82}.
\begin{thm*}[\cite{WZ82}, \cite{Di82}]
An unknown quantum state can't be duplicated.
\end{thm*}
Indeed the cloning process would be given by operator
$$
\psi\otimes(\mbox{state of the Cloning Machine})\mapsto
\psi\otimes\psi\otimes (\mbox{another state of the Machine})$$
which is {\it quadratic} in state vector $\psi$ of the quantum
system. \qed

\subsubsection{Inaccessibility of quantum information}
%\subsubsection{}
As another application of linearity consider the following
%\marginpar{\footnotesize
%Add references. May be discuss relation with Kant's
%thing-in-itself (noumenon).}
\begin{thm*}
No information on quantum system can be gained without destruction
of its state.
\end{thm*}
Indeed the measurement process is described by linear operator
%evolution operator in this case should acts as follows
$$U: \psi_{\mathrm{ini}}\otimes\Psi_{\mathrm{ini}}\mapsto \psi_{\mathrm{fin}}\otimes\Psi_{\mathrm{fin}},$$
where $\psi$ and $\Psi$ are states of the system and the
measurement device
%apparatus
respectively.  The initial state $\Psi_{\mathrm{ini}}$ of the
apparatus supposed to be fixed once and for all, so that
 the final state
$\psi_{\mathrm{fin}}\otimes\Psi_{\mathrm{fin}}$ is a linear
function of $\psi_{\mathrm{ini}}$. This is possible only if
\begin{itemize}\item $\psi_{\mathrm{fin}}$ is linear in
$\psi_{\mathrm{ini}}$ and $\Psi_{\mathrm{fin}}$ is independent of
$\psi_{\mathrm{ini}}$,
%i.e. final state of the apparatus contains no information about the system;
\item or vice versa $\Psi_{\mathrm{fin}}$ is linear in
$\psi_{\mathrm{ini}}$ and $\psi_{\mathrm{fin}}$ is independent of
$\psi_{\mathrm{ini}}$.
\end{itemize}
In the former case the final state of the measurement device
contains no information on the system, while in the latter the
unknown initial state $\psi_{ini}$ is completely erased in the
measurement process. \qed

Emmanuel Kant, who persistently defended absolute reality of
unobservable ``thing-in-itself'', or {\it noumenon}, as opposed to
{\it phenomenon}, should be very pleased with this theorem
identifying noumenon with quantum state.
%\begin{rem}

The theorem
%implies
suggests that complete separation of a system from a measuring
apparatus is
%highly
unlikely. As a rule the system
remains {\it entangled,} with the measuring device, with two
exceptions described above.
%when it gains { full} information
%on the system, or no information at all.
%\end{rem}

%Notice that entanglement of the quantum system with measurement
%apparatus means that the system {\it per se} is in a mixed state
%with density operator
%$$\rho_{sys}=\psi^*\psi$$
%where we identify tensor $\psi=\sum_{ij}\psi^{ij}e_i\otimes f_j\in
%{ \mathcal{H}}_{sys}\otimes{\mathcal{H}}_{app}$ with linear
%operator $\psi:{\mathcal{H}}_{sys}\rightarrow
%{\mathcal{H}}_{app}$, $a\mapsto
%\langle\psi\mid a\rangle= \sum_{ij}\overline{\psi^{ij}}\langle e_i,
%a\rangle b_j$.

\subsection{Reduced states and first glimpse of
entanglement}\label{reduced} Density matrix of composite system
$AB$ can be written as a linear combination of separable states
\begin{equation}\label{sep_decomp}
\rho_{AB}=\sum_\alpha a_\alpha \rho_A^\alpha\otimes \rho_B^\alpha,
\end{equation}
where $\rho_A^\alpha, \rho_B^\alpha$ are mixed states of the
components $A,B$ respectively, and the coefficients $a_\alpha$ are
{\it not\/} necessarily positive.
%$\mathcal{H}_A,\mathcal{H}_B$ respectively.
Its {\it reduced matrices\/} or {\em marginal states\/} may be
defined by equations
\begin{eqnarray*}
\rho_A&=\sum_\alpha
a_\alpha\mathrm{Tr\,}(\rho_B^\alpha)\rho_A^\alpha:=\mathrm{Tr\,}_B(\rho_{AB}),\\
\rho_B&=\sum_\alpha
a_\alpha\mathrm{Tr\,}(\rho_A^\alpha)\rho_B^\alpha:=\mathrm{Tr\,}_A(\rho_{AB}).
\end{eqnarray*}
%Marginal state $\rho_A$
The reduced states $\rho_A,\rho_B$ are independent of the
decomposition (\ref{sep_decomp}) and  can be characterized
intrinsically by the following property
\begin{equation}
\langle X_A\rangle_{\rho_{AB}}=\mathrm{Tr\,}(\rho_{AB}X_A)=\mathrm{Tr\,}(\rho_A
X_A)=\langle X_A\rangle_{\rho_A},\quad\forall\quad
X_A:\mathcal{H}_A\rightarrow\mathcal{H}_A,
\end{equation}
which tells that $\rho_A$ is a ``visible'' state of subsystem $A$.
This justifies the termino\-logy.

\begin{Ex}\label{Schmidt_ex} Let's identify pure state of two component system
$$\psi=\sum_{ij}\psi_{ij}\;\alpha_i\otimes\beta_j\in\mathcal{H}_A\otimes\mathcal{H}_B$$
with its matrix $[\psi_{ij}]$ in orthonormal bases
$\alpha_i,\beta_j$ of $\mathcal{H}_A,\mathcal{H}_B$. Then the
reduced states of $\psi$ in respective bases are given by matrices
\begin{equation}\label{2-comp}\rho_A=\psi^\dag\psi,\quad\rho_B=\psi\psi^\dag,
\end{equation}
which have the same non negative spectra
\begin{equation}\label{isospect}
\mathrm{Spec\,}\rho_A=\mathrm{Spec\,}\rho_B=\lambda
\end{equation}
except extra zeros if $\dim\mathcal{H}_A\ne\dim\mathcal{H}_B$. The
isospectrality implies so called {\it Schmidt decomposition\/}
\begin{equation}\label{Schmidt}
\psi=\sum_i\sqrt{\lambda_i}\;\psi^A_i\otimes\psi^B_i,
\end{equation} where
$\psi^A_i,\psi^B_i$ are eigenvectors of $\rho_A,\rho_B$ with the
same eigenvalue $\lambda_i$.

In striking
 %difference with
contrast to the classical case marginals of a pure state
$\psi\neq\psi_A\otimes\psi_B$ are mixed ones,
%(provided the state is indecomposable $\psi\neq\psi_A\otimes\psi_B$).
i.e. as Sr\"odinger put it {\it``maximal knowledge of the whole
does not necessarily  includes the maximal knowledge of its
parts''} \cite{Sch35}. He coined the term {\it entanglement\/}
just to describe this phenomenon.
%Schr\"odinger \cite{Sch35}
%invented
%introduced the term {\it entanglement\/} to describe this
%quantum
%phenomenon
%situation
%when ``maximal knowledge of the whole does not includes the
%maximal knowledge of its parts''.
Von Neumann entropy of the
marginal states provides a natural measure of entanglement
\begin{equation}\label{entropy}
E(\psi)=-\mathrm{Tr\,}(\rho_A\log\rho_A)=-\mathrm{Tr\,}(\rho_B\log\rho_B)=-\sum_i\lambda_i\log\lambda_i.
\end{equation}
\end{Ex}
In equidimensional system $\dim\mathcal{H}_A=\dim\mathcal{H}_B=n$
maximum of entanglement, equal to $\log n$ entangled bits (ebits),
is attained for a state with scalar reduced matrices
$\rho_A,\rho_B$.
%\begin{Ex*} \label{slice}
%A similar description holds  for multicomponent systems. For
%example, write orthonormal components of tensor
%$\psi\in\mathcal{H}_A\otimes\mathcal{H}_B\otimes\mathcal{H}_C$
%into a cubic matrix $[\psi_{ijk}]$. Then univariant margins of
%$\psi$ are given by {\it Gram matrices\/} formed by Hermitian dot
%products of parallel slices of $[\psi_{ijk}]$.
%%This implies
%It follows that $\rk\rho_C\le \rk\rho_A\cdot\rk\rho_B$, because
%$\rho_C$
%%can be written as
%is Gram matrix of the slices of dimension
%$\rk\rho_A\cdot\rk\rho_B$. This inequality is a simplest
%manifestation of general problem about relations between margins
%of a pure state, which we address below.
%\end{Ex*}

%\begin{rem*} In tensor algebra the above reduction
%%marginalization
%$\rho_{AB}\mapsto\rho_A$ is known as {\it contraction\/}. Most
%mathematicians are familiar with this procedure from differential
%geometry, where, say, Ricci curvature
%$\operatorname{Ric}:\mathcal{T}\rightarrow\mathcal{T}$ is defined
%as contraction of Riemann curvature
%$R:\mathcal{T}\otimes\mathcal{T}\rightarrow\mathcal{T}\otimes\mathcal{T}$
%(here  $\mathcal{T}$ stands for tangent bundle). The results we'll
%discuss below impose some restrictions on possible spectra of
%Riemann and Ricci curvatures.
%\end{rem*}

%\subsection{} The notion of reduced state make sense in more
%general settings of {quantum dynamical system\/}.

%We define the later as Hilbert space $\mathcal{H}$ together with
%Lie algebra of essential observales

\subsection{Quantum dynamical systems}\label{dynamic}
In the above discussion we tacitly suppose, following von Neumann,
that all observable $X_A:\mathcal{H}_A\rightarrow\mathcal{H}_A$ or
what is the same all unitary transformations
$e^{itX_A}:\mathcal{H}_A\rightarrow\mathcal{H}_A$
%\marginpar
%{\footnotesize Explain this, see notes after  Example \ref{spin1}.}
are equally accessible for manipulation with quantum states.
However physical nature of the system may impose unavoidable
constraints.
%usually
%imposes some restrictions on possible manipulation with quantum
%states.

\begin{Ex}\label{two_comp}
%\marginpar{\footnotesize The components may be even space like
%separated with no classical communication at the moment of
%measurement.}
The components of composite system
$\mathcal{H}_{AB}=\mathcal{H}_A\otimes\mathcal{H}_B$ may be
spatially separated by tens of kilometers,  as in EPR pairs used
in quantum cryptography. In such circumstances  only local
observations $X_A$ and $X_B$ are available. This may be even more
compelling if the components are spacelike separated  at the
moment of measurement.
\end{Ex}
\begin{Ex}\label{bose_fermi}
Consider a system of $N$ identical particles, each  with space of
internal degrees of freedom $\mathcal{H}$. By Pauli principle
%imposes strong constraints onto accessible quantum states the
state space of such system shrinks to {\it symmetric tensors\/}
$S^N\mathcal{H}\subset\mathcal{H}^{\otimes N}$ for bosons, and to
{\it skew symmetric tensors\/}
$\wedge^N\mathcal{H}\subset\mathcal{H}^{\otimes N}$  for fermions.
This superselection rule imposes severe restricion on manipulation
with quantum states, effectively reducing
% Because of
%indistinguishability of the particles
the accessible measurements
%observables
%upon
%for such system are the same as for
to that of a single particle.
\end{Ex}
%\marginpar
%{\footnotesize Add fermionic and bosonic systems as most instructive examples}
\begin{Ex}\label{spin} State space $\mathcal{H}_s$ of spin $s$ system has dimension
$2s+1$. %Standard
Measurements upon such system are usually confined to spin
projection onto a chosen direction.
%operators
%manipulations with spin states are
%bounded to measure of spin projection $J_x$, $J_y$, $J_z$ which
They generate Lie algebra $\frak{su\,}(2)$ rather then full
algebra of traceless operators $\frak{su\,}(2s+1)$.
\end{Ex}
%Such kind of examples lead
%many researchers
%This observation leads to an important amendment to the von
%Neumann picture \cite{Her, Emch}
%%expressed
%vigorously stated by Robert Hermann:
This consideration led many researchers to the conclusion, that
available observables  should be included in description of any
quantum system from the outset \cite{Her, Emch}. Robert Hermann
stated this thesis as follows:
%\marginpar{\footnotesize Apparently the term dynamical symmetry group was
%first applied to hydrogen atom, where components of angular moment
%and Runge-Lenz vector form orthogonal Lie algebra $\frak{so\,}(4)$
%w.r. to Poisson bracket see \cite{GreiMull94}.}
%as follows:
\begin{quote}
{\it``The basic principles of quantum mechanics seem to require
the postulation of a Lie algebra of observables and a
representation of this algebra by skew-Hermitian operators.''}
% R. Hermann, Lie Groups for physicists, New York, Benjamin, 1966
\end{quote}

We'll refer to the Lie algebra $\frak{L}$ as {\it algebra of
obsevables\/} and to the corresponding group $G=\exp(i\frak{L})$
as {\it dynamical symmetry group\/} of the quantum system in
question.  Its state space $\mathcal{H}$ together with unitary
representation of the dynamical group $G:\mathcal{H}$ is said to
be {\it quantum dynamical system}. In contrast to R.~Hermann we
treat  $\frak{L}$ as algebra of {\it Hermitian,\/} rather then by
skew-Hermitian operators, and include imaginary unit $i$ in the
definition of Lie bracket $[X,Y]=i(XY-YX)$.

The choice of the algebra $\frak{L}$ depends on the {\it
measurements\/} we are able  to perform over the system, or what
is the same  the {\it Hamiltonians\/} which are accessible for
manipulation with quantum states.
%\marginpar{\footnotesize Add Fock
%formalism of creation and annihilation operators for bosons and
%fermions, with dynamical groups $\mathrm{Sp\,}(2n)$ or
%$\mathrm{SO\,}(2n)$ respectively (reducing in both cases to
%$\mathrm{SU\,}(n)$ if the number of particles is fixes). Coherent
%states in fermionic systems were studied by Cartan, see
%\cite{Cart66}. He call them {\it pure spimors}. See also
%\cite{Simon97}, \cite{Brow92}, \cite{Schwinger70}, \cite{Perel86}.
%}

For example, if we are restricted to {\it local measurements\/} of
a system consisting of two remote components $A,B$ with full
access to the local degrees of freedom then the dynamical group is
$\mathrm{SU}(\mathcal H_A)\times
\mathrm{SU}(\mathcal H_B)$ acting in $\mathcal H_{AB}=\mathcal
H_A\otimes \mathcal H_B$.

In settings of Example \ref{bose_fermi} suppose that a single
particle is described by dynamical system $G:\mathcal{H}$. Then
ensemble of $N$ identical particles corresponds to dynamical
system $G:S^N\mathcal{H}$ for bosons, and to
$G:\wedge^N\mathcal{H}$ for fermions.

The dynamic group of spin system from Example \ref{spin} is ${\rm
SU}(2)$ in its spin $s$ representation $\mathcal{H}_s$.
%\begin{Ex} The algebra of essential observables of two component
%system $\mathcal{H}_{AB}=\mathcal{H}_A\otimes\mathcal{H}_B$ of
%example \ref{two_comp} consists of operators
%$$\frak{L}=\{X_A\otimes 1+1\otimes X_B\mid X_A\in
%su(\mathcal{H}_A), X_B\in
%su(\mathcal{H}_B)\}=su(\mathcal{H}_A)+su(\mathcal{H}_B)$$ and its
%dynamical group amounts to local unitaries $G={\rm
%SU}(\mathcal{H}_A)\times{\rm SU}(\mathcal{H}_B)$. The dynamical
%symmetry group in example \ref{spin} is spin group ${\rm SU}(2)$
%in its spin $s$ representation $\mathcal{H}_s$.
%\end{Ex}
%\subsection{Quantum logic} Put this section before quantum
%dynamical systems? Quantum logic introduced by Birkhoff and von
%Neumann \cite{Bir_Neu36, Pit89} as linear geometry embedded in
%Schubert calculus in Grassmannian and flag varieties, see
%\cite{Ful97}, \cite{Mac91}. Explain Littlewood--Richardson rule.
%Representation subspaces by coherent states (Slater determinant).
%Entangled system of quantum questions and Mumford criterion for of
%stability of a system of linear subspaces. Naimark theorem on
%decomposition of a scalar as sum of nonnegative operators. The
%same for arbitrary observables.

\section{Coherent states}
Coherent states, first introduced by Schr\"odinger \cite{Sch26} in
1926, lapsed into obscurity for decades until Glauber \cite{Glaub}
recovered them in 1963 in connection with laser emission. He have
to wait more then 40 years to win Nobel Prize in 2005 for three
paper published in 1963-64.

Later in 70th Perelomov \cite{Perel72,Perel86} puts coherent
states into
%an adequate context
general framework of dynamic symmetry groups. We'll use a similar
approach for entanglement, and to warm up recall here some basic
facts about coherent states.

\subsection{Glauber coherent states}\label{q_oscil} Let's start with {\it quantum oscillator}, described by canonical
pair of operators $p$, $q$,  $[p,q]=i\hbar$, generating  {\it
Weyl-Heisenberg algebra} $\mathcal W$. This algebra has unique
unitary irreducible representation, which can be realized in {\it
Fock space} $\mathcal F$ spanned by orthonormal set of
$n$-excitations states $|n\rangle$ on
 which dimensionless annihilation and creation operators
$$a=\frac{q+ip}{\sqrt{2\hbar}},\quad
a^\dag=\frac{q-ip}{\sqrt{2\hbar}}, \quad [a,a^\dag]=1$$ act by
formulae
$$a|n\rangle=\sqrt{n}|n-1\rangle, \quad
a^\dag|n\rangle=\sqrt{n+1}|n+1\rangle.$$

A typical element from {\it Weyl-Heisenberg group\/}
$W=\exp\mathcal W$, up to a phase factor, is of the form
$D(\alpha)=\exp(\alpha a^\dag -\alpha^* a)$ for some
$\alpha\in\Bbb C$. Action of this operator on vacuum $|0\rangle$
produces state
\begin{equation}\label{Glauber}|\alpha\rangle:=D(\alpha)|0\rangle=
\exp\left(-\frac{|\alpha|^2}{2}\right)
\sum_{n\ge 0}\frac{\alpha^n}{\sqrt{n!}}|n\rangle,
\end{equation}
known as {\it Glauber coherent state}. The number of excitations
in this state has Poisson distribution with parameter
$|\alpha|^2$. In many respects its behavior  is close to
classical, e.g. Heisenberg's uncertainty $\Delta p\Delta
q=\hbar/2$ for this state is minimal possible. In coordinate
representation
$$q=x,\quad p=i\hbar\frac{d}{dx}$$
its time evolution is given by harmonic oscillation of Gaussian
distribution of width $\sqrt{\hbar}$ with amplitude
$|\alpha|\sqrt{2\hbar}$. Therefor for big number of photons
$|\alpha|^2\gg 1$ coherent states behave classically. Recall also
Glauber's theorem \cite{Glaub64} which claims that classical field
or force excites quantum oscillator into a coherent state.
% see
%R.~Glauber, {\it Optical coherence and the photon statistics}, in
%Quantum optics and electronics, C.~DeWitt, A.~Blandin,
%C.~Cohen-Tannoudji, ed., Gorgon and Breach, New York, 1965.

We'll return to these  aspects of coherent states later, and focus
now on their mathematical description
%We can summarize this as follows:
\begin{quote}\centerline{\it Glauber coherent states =  $W$-orbit of
vacuum}\end{quote} which sounds more suggestive then explicit
equation (\ref{Glauber}).
%\end{displaymath}

\subsection{General coherent
states}

%To turn this property into a precise definition,
Let's now turn to arbitrary quantum system $A$ with dynamical
symmetry group $G=\exp i\frak{L}$. By definition its  Lie algebra
$\frak{L}={\rm Lie\,} G$ is generated by all {\it essential
observables\,} of the system (like $p,q$ in the above example). To
simplify the underling mathematics suppose in addition that state
space of the system $\mathcal{H}_A$ is finite, and representation
of $G$ in $\mathcal{ H}_A$ is irreducible.

To extend (\ref{Glauber}) to this general setting  we have to
understand the special role of the vacuum, which primary
considered as a {\it ground state} of a system. For
group-theoretical approach, however, another its property is more
relevant:
 \begin{quote}\centerline{\it Vacuum is a state with
maximal symmetry\/.}\end{quote} This may be also spelled out that
vacuum is a most degenerate state of a system.
\subsection{Complexified dynamical group}
Symmetries of state $\psi$ are given by its {\it stabilizers}
\begin{equation}\label{stabilizer} G_\psi=\{g\in G\mid g\psi=\mu(g)\psi\},\quad {\frak
L}_\psi=\{X\in{\frak L}\mid X\psi=\lambda(X)\psi\}
\end{equation} in the
dynamical group $G$ or in its Lie algebra ${\frak L}={\rm
Lie\,}G$. Here $\mu(g)$ and $\lambda(X)$ are scalars. Looking back
to the quantum oscillator, we see that some symmetries are
actually hidden, and manifest themselves only in {\it
complexified} algebra ${\frak L}^c={\frak L}\otimes{\mathbb C}$
and group $G^c=\exp{\frak L}^c$. For example, stabilizer of vacuum
$|0\rangle$ in Weyl-Heisenberg algebra $\mathcal W$ is trivial
$\mathcal{W}_{|0\rangle}=\mbox{scalars}$, while in complexified
algebra ${\mathcal W}^c$ it contains annihilation operator,
${\mathcal W}_{|0\rangle}^c={\mathbb C}+{\mathbb C} a$. In the
last case the stabilizer is big enough to recover the whole
dynamical algebra
$${\mathcal W}^c={\mathcal W}^c_{|0\rangle}+{{\mathcal W}^c_{|0\rangle}}^\dag.$$
This decomposition, called {\it complex polarization}, gives a
precise meaning for the maximal degeneracy of a vacuum or a
coherent state.
%It ensures that dimension of the symmetry group of
%such state is at least half of dimension the whole dynamical group.
\subsection{General definition of coherent state}\label{gen_coh}
State $\psi\in \mathcal{H}$ is said to be {\it coherent} if
$$%\framebox{\parbox{2.5cm}{\begin{center}$
{\mathfrak L}^c={\frak L}^c_\psi+{{\mathfrak L}^c_\psi}^\dag
%$\end{center}}}
$$
In finite dimensional case all such decompositions come from {\it
Borel subalgebra}, i.e. a maximal solvable subalgebra ${\frak
B}\subset{\frak L}^c$. The corresponding {\it Borel subgroup\/}
$B=\exp\frak{B}$ is a minimal subgroup of $G^c$ with compact
factor $G^c/B$.
%called {\it flag variety\/} of the group.
Typical example is subgroup of upper triangular matrices
%${\frak T}\subset{\rm Mat}(n,{\mathbb C})$
in $\mathrm{SL\,}(n,\mathbb{C})$ = complexification of
$\mathrm{SU\,}(n)$. It is a basic structural fact that ${\frak
B}+{\frak B}^\dag={\frak L}^c$, and therefore
$$%\framebox
{\parbox{7cm}{\begin{center}$\psi$ is coherent
$\Leftrightarrow\psi$ is an eigenvector of  $\frak
B$\end{center}}}$$ In representation theory eigenstate $\psi$ of
${\frak B}$  is called {\it highest vector}, and the corresponding
eigenvalue $\lambda=\lambda(X)$,
$$X\psi=\lambda(X)\psi,\quad X\in {\frak B}$$
is said to be {\it highest weight}.

Here are the basic properties of coherent states.

\begin{itemize}
\item
For irreducible system $G:{\mathcal{H}}$ the highest vector
$\psi_0$ (=vacuum) is unique.
\item
There is only one irreducible representation ${\mathcal{
H}}={\mathcal H}_\lambda$ with highest weight $\lambda$.
\item
All coherent states are of the form $\psi=g\psi_0$, $g\in G$.
\item
Coherent state $\psi$ in composite system ${\mathcal
H}_{AB}={\mathcal H}_A\otimes{\mathcal H}_B$ with dynamical group
$G_{AB}=G_A\times G_B$ splits into product $\psi=\psi_1\otimes
\psi_2$ of coherent states of the components.
\end{itemize}
%One can spell out these properties by saying that unitary
%irreducible representations of group $G$ are parameterized by
%symmetry type of their coherent states or vacua.
\begin{rem*}Coherent state theory, in the form given by Perelomov \cite{Perel86}, is a
physical equivalent of Kirillov--Kostant {\it orbit method}
\cite{Kiril76} in representation theory.
\end{rem*}

The complexified group play crucial role in our study. Its
operational interpretation  may vary. Here is a couple of
examples.
\begin{Ex}\label{spin1} {\it Spin systems.} For system of spin $s$ (see example
\ref{spin}) coherent states have definite
%maximal possible
spin projection $s$ onto some direction
$$%\framebox{\parbox{4.5cm}{\begin{center}$
\psi \mbox{ is coherent } \Longleftrightarrow
\psi=|s\rangle.
%\end{center}}}
$$
%In many cases
%The complexified symmetry group is physically meaningful. For a
%system of spin $s>0$
Complexification of
%the dynamical
spin group ${\rm SU}(2)$
%Its complexification
is group of unimodular matrices ${\rm SL}(2,{\mathbb C})$.
% which
%, as first noted by Wigner in 1939,
The latter is
%a double cover of
locally isomorphic to {\it Lorentz group\/} and
%it is responsible for
controls relativistic transformation of spin states in a moving
frame.
\end{Ex}

%Group of {\it complex} symmetries of a coherent state is conjugate
%to group of triangular matrices, while ${\rm SU}(2)$ symmetries
%amounts to diagonal matrices.
%\end{Ex}
\begin{Ex}\label{SLOCC} For two
component system $\mathcal{H}_{AB}=
\mathcal{H}_A\otimes\mathcal{H}_B$ with full access to local
degrees of freedom the coherent states are decomposable ones
$$\psi_{AB} \mbox{ is coherent}\Longleftrightarrow
\psi_{AB}=\psi_A\otimes\psi_B.$$
The dynamical group of this system is
%$$\frak{L}=\{X_A\otimes1+1\otimes X_B\},\quad
$G={\rm SU}(\mathcal{H}_A)\times{\rm SU}(\mathcal{H}_B)$, see
example \ref{two_comp}.
%consists of observables $X_A\otimes1+1\otimes X_B$,
%\begin{eqnarray*}G&=&{\rm SU}(\mathcal{H}_A)\times{\rm SU}(\mathcal{H}_B),\\
%G_\mathbb{C}&=&{\rm SL}(\mathcal{H}_A)\times{\rm SL}(\mathcal{H}_B).
%\end{eqnarray*}
Its complexification $G^c={\rm SL}(\mathcal{H}_A)\times{\rm
SL}(\mathcal{H}_B)$  has an important quantum informational
interpretation as group of invertible Stochastic Local Operations
assisted with Classical Communication (SLOCC transformations), see
\cite{VDMV}. These are essentialy LOCC operations with
postselection.
\end{Ex}
\subsection{Total variance}
Let's define {\it total variance} of state $\psi$ by equation
\begin{equation}\label{Tot_Var}
%\framebox{\parbox{6cm}{\begin{center} $
{\mathbb D}(\psi)=\sum_i
\langle\psi|X_i^2|\psi\rangle-\langle\psi|X_i|\psi\rangle^2
%$\end{center}}}
\end{equation} where
%the sum runs over
$X_i\in\frak{L}$ form an orthonormal basis  of the Lie algebra of
essential observables  with respect to its invariant metric (for
spin group ${\rm SU}(2)$ one can take for the basis spin projector
operators $J_x$, $J_y$, $J_z$). The total variance is independent
of the basis $X_i$, hence $G$-invariant.
% $\mathbb D(g\psi)=\mathbb D(\psi)$, $ g\in G$.
 {It measures the total level of {\it quantum
fluctuations\/}} of the system in state $\psi$.

The first sum in (\ref{Tot_Var}) contains well known {\it Casimir
operator}
$$C=\sum_i X_i^2$$
which commutes with $G$ and hence acts as a scalar in every
irreducible representation. Specifically
%\marginpar{\footnotesize
%State this as a theorem and explain why $C$ is independent of the
%basis.}
\begin{thm}
%Let $G:\mathcal{H}_\lambda$ be irreducible
%representation with highest weight $\lambda$. Then
The Casimir operator $C$ acts in irreducible representation
$\mathcal{H}_\lambda$ of highest weight $\lambda$ as
multiplication by scalar
$C_\lambda=\langle\lambda,\lambda+2\delta\rangle$.
\end{thm}

{\footnotesize
%More generally
One can use two {\it dual bases} $X_i$ and $X^j$ of $\frak{L}$,
with respect to invariant bilinear form $B(X_i,X^j)=\delta_{ij}$
to construct the Casimir operator
$$C=\sum_iX_iX^i.$$
For example, take basis of $\frak{L}$ consisting of orthonormal
basis $H_i$ of Cartan subalgebra $\frak{h}\subset\frak{L}$ and its
{\it root vectors} $X_\alpha\in\frak{L}$ normalized by condition
$B(X_\alpha,X_{-\alpha})=1$. Then the dual basis is obtained by
substitution $X_\alpha\mapsto X_{-\alpha}$ and hence
$$
C=\sum_iH_i^2+\sum_{\alpha=\mathrm{root}}X_\alpha
X_{-\alpha}=\sum_iH_i^2+\sum_{\alpha>0}
H_\alpha+2\sum_{\alpha>0}X_{-\alpha}X_{\alpha},
$$
where in the
last equation we use commutation relation
$[X_\alpha,X_{-\alpha}]=H_\alpha$. Applying this to the highest
vector $\psi\in\mathcal{H}$ of weight $\lambda$, which by
definition is annihilated by all operators $X_\alpha$, $\alpha>0$
 and $H\psi=\lambda(H)\psi$, $H\in\frak{h}$, we get
\begin{equation}\label{Casimir_eigen}
C\psi=\sum_i\lambda(H_i)^2\psi+\sum_{\alpha>0}\lambda(H_\alpha)\psi=
\langle\lambda,\lambda+2\delta\rangle\psi,
\end{equation}
where $2\delta=\sum_{\alpha>0}\alpha$ is the sum of positive roots
and $\langle*,*\rangle$ is the invariant form $B$ translated to
the dual space $\frak{h}^*$. Hence Casimir operator $C$ acts as
scalar $C_\lambda=\langle\lambda,\lambda+2\delta\rangle$ in
irreducible representation with highest weight $\lambda$.\qed}

\subsection{Extremal property of coherent states}\label{extr_coh}

For spin $s$ representation ${\mathcal H}_s$ of ${\rm SU}(2)$ the
Casimir is equal to square of the total moment
$$C=J^2=J_x^2+J_y^2+J_z^2=s(s+1).$$
%In general
%$C=\langle\lambda,\lambda+2\delta\rangle$ for irreducible
%representation ${\mathcal H}_\lambda$ with highest weight
%$\lambda$ (here we use H.~Weyl's notation $\delta$ for halfsum of
%positive roots).
Hence
\begin{equation}\label{c_var}\mathbb
D(\psi)=\langle\lambda,\lambda+2\delta\rangle-
\sum_i\langle\psi|X_i|\psi\rangle^2.
\end{equation}

\begin{thm}[Delbourgo and Fox \cite{Del}]\label{Del-Fox} State $\psi$ is coherent iff its
total variance is minimal possible, and in this case
$${\mathbb D}(\psi)=\langle\lambda,2\delta\rangle.$$
\end{thm}

{\footnotesize Let $\rho=|\psi\rangle\langle\psi|$ be pure state
and $\rho_{\frak{L}}$ be its orthogonal projection into subalgebra
$\frak{L}\subset\mathrm{Herm}(\mathcal{H})$ of algebra of all
Hermitian operators in $\mathcal{H}$ with  trace metric
$(X,Y)=\mathrm{Tr}(X\cdot Y)$. By definition we have
$$\langle\psi|X|\psi\rangle=\mathrm{Tr}_\mathcal{H}(\rho
X)=\mathrm{Tr}_\mathcal{H}(\rho_\frak{L}X),\quad\forall
X\in\mathfrak{L}.$$ Choose a Cartan subalgebra
$\frak{h}\subset\frak{L}$ containing $\rho_\frak{L}$. Then
$\langle\psi|X_i|\psi\rangle=\mathrm{Tr}_\mathcal{H}(\rho_\frak{L}X_i)=0$
for $X_i\bot\frak{h}$ and we can restrict the sum in (\ref{c_var})
to orthonormal basis $H_i$ of Cartan subalgebra
$\frak{h}\subset\frak{L}$ for which by the definition of highest
weight $\langle\psi|H|\psi\rangle^2\le \lambda(H)^2$ with equality
for the highest vector $\psi$ only. Hence
\begin{equation}\label{coh_ineq}
\sum_i\langle\psi|X_i|\psi\rangle^2=
\sum_i\langle\psi|H_i|\psi\rangle^2\le\sum_i\lambda(H_i)^2=
\langle\lambda,\lambda\rangle,
\end{equation}
and therefore $\mathbb{D}(\psi)\ge
\langle\lambda,\lambda+2\delta\rangle-\langle\lambda,\lambda\rangle=\langle\lambda,2\delta\rangle$,
with equality for coherent states only. \qed

}

%The theorem belongs to Delbourgo and Fox \cite{Del}.
The theorem supports the thesis
%a common belief,
that coherent states are closest to classical ones, cf. $n^\circ$
\ref{q_oscil}. Note however that such simple characterization
holds only for finite dimensional systems. The total variance, for
example, makes no sense for quantum oscillator, for which we have
{\it minimal uncertainty} $\Delta p\Delta q=\hbar/2$ instead.

\begin{Ex} For coherent state of spin $s$ system Theorem \ref{Del-Fox} gives
$\mathbb{D}(\psi)=s$. Hence amplitude of quantum fluctuations
$\sqrt{s}$ for such state is of smaller order then spin $s$, which
by Example \ref{spin1} has a definite direction. Therefor for
$s\rightarrow\infty$ such state
%behaves as
looks like a classical rigid body rotating around the spin axis.
%classically.
% Recall from
%example \ref{spin1} that for spin $s$
%%representation of ${\rm SU}(2):\mathcal{H}_s$
%system coherent state $\psi$ has definite spin projection $s$ onto
%some direction.
%%and
%%By the theorem
%%The standard deviation
%The amplitude of quantum fluctuations $\sqrt{{\mathbb
%D}(\psi)}=\sqrt{s}$
%%. The standard deviation
%%of quantum fluctuations
%%$\sqrt{s}$
% for such state
%%in this case
%is of smaller order then $s$, therefore for
%$s\rightarrow
%\infty$ it behaves classically.
\end{Ex}

\subsection{Quadratic equations defining coherent
states}\label{quad_coh} There is another useful description of
coherent states by a system of quadratic equations.
\begin{Ex} Consider two component system
$\mathcal{H}_{AB}=\mathcal{H}_A\otimes\mathcal{H}_B$ with full
access to local degrees of freedom
$G=\mathrm{SU\,}(\mathcal{H}_A)\otimes\mathrm{SU\,}(\mathcal{H}_B)$.
Coherent states in this case are just separable states
$\psi=\psi_A\otimes\psi_B$ with density matrix
$\rho=|\psi\rangle\langle\psi|$ of rank one. Such matrices can be
characterized by vanishing of all minors of order two. Hence
coherent states of two component system can be described by a
system of quadratic equations.
\end{Ex}
It turns out that a similar description holds for arbitrary
irreducible system $G:\mathcal{H}_\lambda$ with highest weight
$\lambda$, see \cite{Licht82}.
\begin{thm}\label{Licht}
State $\psi\in \mathcal{H}_\lambda$  is coherent iff
$\psi\otimes\psi$ is eigenvector of the Casimir operator $C$ with
eigenvalue $\langle2\lambda+2\delta,2\lambda\rangle$
%in the
%doublet $G:\mathcal{H}_\lambda\otimes
%\mathcal{H}_\lambda$ the following equation holds
\begin{equation}\label{quad_coh_eqn}
C(\psi\otimes\psi)=\langle 2\lambda+2\delta,2\lambda\rangle
(\psi\otimes\psi).
\end{equation}
\end{thm}
{\footnotesize Indeed, if $\psi$ is highest vector of weight
$\lambda$ then $\psi\otimes\psi$ is a highest vector of weight
$2\lambda$ and equation (\ref{quad_coh_eqn}) follows from
%the eigenvalue of Casimir
(\ref{Casimir_eigen}).

Vice versa, in terms of orthonormal basis $X_i$ of Lie algebra
$\frak{L}=\mathrm{Lie\,} G$ the Casimir operator in the doublet
$\mathcal{H}_\lambda\otimes
\mathcal{H}_\lambda$ looks as follows
$$C=\sum_i(X_i\otimes1+1\otimes X_i)^2=\sum_i
X_i^2\otimes1+1\otimes X_i^2+2\sum_iX_i\otimes X_i.$$ Hence under
conditions of the theorem
$$\langle2\lambda+2\delta,2\lambda\rangle=\langle\psi\otimes\psi|C|\psi\otimes
\psi\rangle=2\langle\lambda+2\delta,\lambda\rangle+2\sum_i\langle\psi|X_i|\psi\rangle^2.$$
It follows that
$$\sum_i\langle\psi|X_i|\psi\rangle^2=\langle\lambda,\lambda\rangle$$
and hence by inequality (\ref{coh_ineq}) state $\psi$ is
coherent.\qed }
\begin{rem}
 The above calculation show that equation
(\ref{quad_coh_eqn}) is equivalent to
%\marginpar{\footnotesize
%For Casimir equal to Laplace operator this resembles eikonal
%equation.}
\begin{equation}\label{quad_coh_eqn2}
\sum_iX_i\psi\otimes
X_i\psi=\langle\lambda,\lambda\rangle\;\psi\otimes\psi,
\end{equation}
which in turn amounts to a system of {\it quadratic equations\/}
on the components of a coherent state $\psi$.
\end{rem}

%Equation (\ref{quad_coh_eqn}) amounts to a system of quadratic
%equations on components of the state $\psi$.
\begin{Ex}
For spin $s$ system
%equation (\ref{quad_coh_eqn})
the theorem tells that state $\psi$ is coherent iff
$\psi\otimes\psi$ has definite spin $2s$.  Equations
(\ref{quad_coh_eqn2}) amounts to
$$J_x\psi\otimes J_x\psi+J_y\psi\otimes J_y\psi+J_z\psi\otimes
J_z\psi=s^2\psi\otimes\psi.$$
%As an example let's consider spin $s$ states
%$$\psi=\sum_{\mu=-s}^{s} a_{\mu}{2s\choose \mu+s}|\mu\rangle.$$
\end{Ex}

%\marginpar
%{\footnotesize There is another characterization of the cone of
%coherent states by quadratic equations
%$C(\psi\otimes\psi)=\langle2\lambda,2\lambda+2\delta\rangle(\psi\otimes\psi)$,
%see \cite{Licht82}. This is a natural extension of the quadratic
%relations between Pl\"ucker coordinates (=Slater determinants) in
%Grassmannian. For two component system they amounts to vanishing
%of the minors of order two of the matrix of separable state
%$\psi\in\mathcal{H}_A\otimes\mathcal{H}_B$. Add here further
%analysis of the variance from {\tt Baris-APL.tex} with application
%to generalized concurrence. May be this is the right place to
%mention Berenstein-Sjamaar theorem \cite{Ber-Sja}. }

\section{Entanglement}

%\subsection{}
From a thought experiment for testing the very basic principles of
quantum mechanics in its earlier years \cite{EPR, Sch35}
entanglement nowadays is growing into an important tool for
quantum information processing. Surprisingly enough so far there
is no agreement among the experts on  the very definition and the
origin of entanglement, except unanimous conviction in its
fundamental nature and in necessity of its better understanding.

Here we
%propose
discuss a novel  approach to entanglement \cite{Kl02}, based on
dynamical symmetry group, which puts it into a broader context,
eventually applicable to all quantum systems. This sheds new light
on known results providing for them a unified conceptual
framework, opens a new prospect for further development of the
subject, reveals its deep and unexpected connections with other
branches of physics and mathematics, and
%last but not least
provides an insight on conditions in which entangled state can be
stable.

\subsection{What is entanglement?}
Everybody knows, and nobody understand what is entanglement. Here
are some virtual answers to the question borrowed from Dagmar
Bru{\ss} collection \cite{Bruss}:
\begin{itemize}
\item J.~Bell: $\ldots$ {\it a correlation that is stronger then any
classical correlation.}
\item D.~Mermin: $\ldots$ {\it a correlation that contradicts the
theory of elements of reality.}
\item A.~Peres: $\ldots$ {\it a trick that quantum magicians use to produce phenomena that
cannot be imitated by classical magicians.}
\item  C.~Bennet: $\ldots$ {\it a resource that enables quantum
teleportation.}
\item P.~Shor: $\ldots$ {\it a global structure of wave function that
allows the faster algorithms.}
\item A.~Ekert: $\ldots$ {\it a tool for secure communication.}
\item Horodecki family: $\ldots$ {\it the need for first application of
positive maps in physics.}
\end{itemize}
This list should be enhanced with extensively cited
Schr\"odinger's definition given in $n^\circ$\ref{reduced}.
 The very term
 %``entanglement of our knowledge''
 was coined by Schr\"odinger in the famous
``cat paradox''
%\footnote{As BBC puts it: {\it In quantum mechanic it is not so
%easy to be or not to be.}}
paper \cite{Sch35}
%to describe the situation when ``maximal
%knowledge of the whole does not necessarily includes the maximal
%knowledge of its parts''.  Schr\"odinger paper
which in turn was inspired by the no less celebrated
Einstein--Podolsky--Rosen {\it gedanken} experiment \cite{EPR}.
While the latter authors were amazed by nonlocal nature of
correlations between the involved particles, J.~Bell was the first
to note that the correlations themselves, puting aside the
nonlocality, are inconsistent with classical realism. Since then
Bell's inequalities are produced in industrial quantities and
remain the main tool for testing ``genuine'' entanglement. Note
however that in some cases  LOCC operations can transform
%sate compatible with classical realism
a classical state into nonclassical one \cite{Pop}. Besides in a
sense every quantum system of dimension at least three is
nonclassical, see $n^\circ$\ref{pent} and
\cite{Malley04,Malley05}.
%\cite{Peres}, even if it consists of only one component.

%{\color{red} Rewrite this! Add various definitions from D.~Bru\ss,
%{\it Characterizing entanglement}, J. Math. Phys., 43(2002), No 9,
%4237--4251.}
Below we briefly discuss the nonlocality and violation of
classical realism. Neither of this effects, however, allow
decisively
%distinguish
characterize entangled states.
%from others.
Therefor eventually we turn to another approach, based on the
dynamical symmetry group.

\subsection{EPR paradox}
Decay of a spin zero state into two components of spin 1/2
subjects to a strong correlation between spin projections of the
components, caused by conservation of moment. The correlation
creates an apparent {\it information channel\/} between the
components, acting beyond their light cones.

%It is important to realize
Let me emphasize that quantum mechanics refuted the possibility
that the spin projection have been  fixed at the moment of decay,
rather then at the moment of measurement. Otherwise two spatially
separated observers can see the same event like burst of a
supernova simultaneously even if they are spacelike separated, see
\cite{Per_Ter03}. There is no such ``event'' or ``physical
reality'' in the Bohm version of EPR experiment.

This paradox, recognized in early years of quantum mechanics
\cite{EPR, Bohm}, nowadays has many applications, but no intuitive
explanation. It is so disturbing that sometimes physicists just
%still
ignore it. For example,
%almost every
one of the finest  recent book justifies QFT commutation relations
as follows \cite{Zee}:
%{\footnotesize The emphasis here should be
%on the absence of hidden variables. For example Bell's inequality
%refute the possibility that spin states of the components are
%fixed at the moment of decay. Otherwise two spatially separated
%observers can see the same event like burst of a supernova
%simultaneously even if they are spacelike separated, see
%\cite{Per_Ter03}. There is no such ``event'' or ``physical reality''
%in the Bohm version of EPR experiment.}
\begin{quote}
%A basic quantum principle states that if two observables commute
%then they are simultaneously diagonalizable and hence observable.
{\it A basic relativistic principle states that if two spacetime
points are spacelike with respect to each other then no signal can
propagate between them, and hence the measurement of an observable
at one of the points cannot influence the measurement of another
observable at the other point.}
\end{quote}
Experiments with EPR pairs tell just the opposite
\cite{ADR82,Gen05}. I am not in position to comment this {\it
nonlocality\/} phenomenon, and therefor turn
%confine myself instead
to less involved {\it Bell's approach\/},
%Henceforth we completely disregard the nonlocality, and turn  to
%confined to
limited  to the quantum correlations {\it per se}.

\subsection{Bell's inequalities}\label{Bell}
%and classical marginal problem}
Let's start with {\it classical marginal problem\/} which  asks
for existence of a ``body'' in $\mathbb{R}^n$ with given
projections onto some coordinate subspaces
$\mathbb{R}^I\subset\mathbb{R}^n, I\subset
\{1,2,\ldots,n\}$, i.e. existence of {\it probability density}
$p(x)=p(x_1,x_2,\ldots,x_n)$ with given {\it marginal
distributions}
$$p_I(x_I)=\int_{\mathbb{R}^J}p(x)dx_J, J=\{1,2,\ldots,n\}\backslash I.$$

%From math point of view
In discrete version the classical MP amounts to calculation of an
image of a multidimensional symplex, say
$\Delta=\{p_{ijk}\ge0|\sum p_{ijk}=1\}$, under a linear map like
\begin{eqnarray*}\pi:\mathbb{R}^{\ell m
n}&\rightarrow&\mathbb{R}^{\ell
m}\oplus\mathbb{R}^{ mn}\oplus\mathbb{R}^{n\ell},\\
p_{ijk}&\mapsto&(p_{ij},p_{jk},p_{ki}),
\end{eqnarray*}
$$p_{ij}=\sum_k p_{ijk},\quad p_{jk}=\sum_i p_{ijk},\quad p_{ki}=\sum_j p_{ijk}.$$
The image $\pi(\Delta)$ is convex hull of $\pi(\mbox{Vertices
}\Delta)$.  So {\it the classical MP amounts to calculation of
facets of a convex hull.} In high dimensions this may be a
computational nightmare \cite{FO85, Pit89}.
%\begin{Ex}\label{Gale} Univariant margins are always compatible, e.g. one can
%consider univariant marginals  $x_i$ as independent variables.
%However under additional constraints, say
%%Under additional constraints
%%For restricted density, say $p(x)=0/1$,
%for a ``body'' of constant density, even univariant marginal
%problem becomes nontrivial. For example, {\it Gale-Ryser theorem}
%\cite{Gale} tells that partitions $\lambda,\mu$ are margins of a
%rectangular $0/1$ matrix iff $\lambda\prec\mu^t$. Here marginal
%values
%%$\lambda_1\ge\lambda_2\ge\cdots\ge\lambda_m\ge0$
%arranged in decreasing order are treated as {\it Young diagrams\/}
%%$\lambda$
%%with $i$-th row of length $\lambda_i$
%{
%$$\lambda=(5,4,2,1)=\yng(5,4,2,1)\quad\lambda^t=(4,3,2,2,1)=\yng(4,3,2,2,1)$$}
%$\mu^t$ stands for {\it transposed\/} diagram,  and the {\em
%majorization order\/} $\lambda\prec\nu$ is defined by inequalities
%{\begin{eqnarray*}\lambda_1&\le&\nu_1\\
%\lambda_1+\lambda_2&\le&\nu_1+\nu_2\\
%\lambda_1+\lambda_2+\lambda_3&\le&\nu_1+\nu_2+\nu_3\\
%\cdots&\cdots&\cdots
%\end{eqnarray*}}
%\end{Ex}

\begin{Ex}\label{class_real}{\it Classical realism.} Let $X_i:\mathcal{H}_A\rightarrow \mathcal{H}_A$ be observables of
quantum system $A$. Actual measurement of $X_i$ produces random
quantity $x_i$ with values in ${\rm Spec\,}(X_i)$ and density
$p_i(x_i)$ implicitly determined by expectations
$$\langle f(x_i)\rangle=\langle\psi|f(X_i)|\psi\rangle$$
for all functions $f$ on spectrum ${\rm Spec\,}(X_i)$.
%\end{slide}
%\begin{slide}\hfill{{\theslide}}\newline\large
 For
{\it commuting} observables $X_i,i\in I$ the random variables
$x_i, i\in I$ have {\it joint distribution} $p_I(x_I)$ defined by
similar equation
\begin{equation}\langle
f(x_I)\rangle=\langle\psi|f(X_I)|\psi\rangle,\quad \forall
f.\label{joint}
\end{equation}
{\it Classical realism} postulates existence of a hidden joint
distribution of {\it all variables} $x_i$. This amounts to
compatibility of the marginal distributions (\ref{joint})
%$p_I(x_I)$
for {\it commuting} sets of observables
$X_I$. %Hence
{\it Bell inequalities}, designed to test classical realism,
%are just another incarnation of
%are byproducts of
stem from the classical marginal problem.
% (studied by mathematicians prior WWII).
%Classical marginal problem
%%appears in quantum physics in form of
%is hidden behind {\color{blue}{\it Bell inequalities}} designed to
%test the classical realism.
\end{Ex}

\begin{Ex}  Observations of disjoint components
of two qubit system $\mathcal{H}_A\otimes\mathcal{H}_B$ always
commute. Let $A_i,B_j$ be spin projection operators in sites $A,B$
onto directions $i,j$. Their observed values $a_i,b_j=\pm1$
satisfy inequality
%Operators $A_i$, $B_j$
%commute, and for spin 1/2 with two measurement per site Kellerer's
%cone $\mathcal{K}$ is given by
%%of all nonnegative functions of the form
%$$F(a_1,a_2,b_1,b_2)=f_{11}(a_1,b_1)+f_{12}(a_1,b_2)+f_{21}(a_2,b_1)+f_{22}(a_2,b_2)
%\ge0,$$
%where $a_i,b_j=\pm1$ are eigenvalues of $A_i$ and $B_j$.
%(we
%consider  the simplest case of
%(for spin 1/2 with two measurement per site).
%All the edges of this cone can be obtained from
%Clauser-Horn-Shimony-Holt inequality \cite{CHSH69}
$$a_1b_1+a_2b_1+a_2b_2-a_1b_2+2\ge0.$$
Indeed product of the monomials $\pm a_ib_j$ in LHS is equal to
$-1$. Hence one of the monomials is equal to $+1$ and sum of the
rest is $\ge-3$.

If all the observables have a hidden joint distribution then
taking the expectations we arrive at {\it
Clauser-Horne-Shimony-Holt\/}
%by permutation of particles $a\leftrightarrow b$ and switching
%eigenvalues $a_i\mapsto \pm a_i$, and $b_j\mapsto
%\pm b_j$. So we have essentially one Bell's type
inequality for testing ``classical realism''
\begin{equation}\label{CHSH}\langle\psi|A_1B_1|\psi\rangle+
\langle\psi|A_2B_1|\psi\rangle+
\langle\psi|A_2B_2|\psi\rangle-
\langle\psi|A_1B_2|\psi\rangle+2\ge0.
\end{equation}
All other marginal constraints can be obtained from
%this inequality
it by spin flips $A_i\mapsto \pm A_i$.
%For two qubits with two spin measurements $a_i,b_j$ per site their
%compatibility
%%with the {\it classical realism}
%is given by {16} inequalities obtained from {\it
%Clauser-Horne-Shimony-Holt} inequality
%$$\langle a_1b_1\rangle+\langle a_2b_1\rangle+\langle
%a_2b_2\rangle-\langle a_1b_2\rangle+2\ge 0$$ for spin projection
%operators $a_i,b_j$ in sites $A,B$ respectively. by spin flips
%$a_i\mapsto \pm a_j$  and permutation of the components
%$A\leftrightarrow B$.
\end{Ex}
%\end{slide}
%\begin{slide}\hfill{{\theslide}}\newline \Large
\begin{Ex}   For three qubits with two
measurements per site the marginal constraints amounts to {53856}
independent inequalities, see \cite{Pit-Svoz}.
%This example may
%help to
%%get rid of
%disabuse us from overoptimistic expectations for the {\it quantum
%marginal problem} to be discussed below.
\end{Ex}
%{\footnotesize Emphasize that
Bell's inequalities
%prohibit
make it impossible to model quantum mechanics by classical means.
In particular, there is no way to reduce quantum computation to
classical one.
%computing can't be modelled by classical means.
%This open a possibility that quantum computer may resolve problems
%which classical computer can't handle in principle, e.g. solve an
%algorithmically unsolvable problem.

\subsection{Pentagram inequality} \label{pent}
%\subsection{Bell's inequalities in spin one system}
%\marginpar
%{\footnotesize Rewrite this as Bell
%inequalities for spin 1 system and deduce that in system of
%dimension $\ge3$ every state is nonclassical. Give bi-photon
%interpretation of spin 1 system.}
Here I'll
%consider in details
give an account of nonclassical states in spin 1 system. Its
optical version, called {\it biphoton\/}, is a hot topic both for
theoretical and experimental studies \cite{Serg00, Serg03,Serg04}.
The so-called {\it neutrally polarised\/} state of biphoton
routinely treated as entangled, since a beam splitter can
transform it into a EPR pair of photons. This is the simplest one
component system which manifests entanglement.
%, see $n^\circ$ \ref{comp_ent}-\ref{gen_ent} for precise definition.
%Let me stress that this is not a
%question of mathematical definition, to be given later in
%$n^\circ$ \ref{comp_ent}. For biphton, an optical version of spin
%1 system, such states
%%called {\it neutrally polarized\/} and
%routinely treated as entangled by experimenters \cite{Serg00}.

%It is convenient to
Spin 1 state space may be identified with complexification of
Euclidean space $\mathbb{E}^3$
$$\mathcal{H}=\mathbb{E}^3\otimes\mathbb{C},$$
where spin group $\mathrm{SU\,}(2)$, locally isomorphic to
$\mathrm{SO\,}(3)$, acts via rotations of $\mathbb{E}^3$. Hilbert
space $\mathcal{H}$ inherits from $\mathbb{E}^3$ bilinear scalar
and cross products,
%that will be
to be denoted by $(x,y)$ and $[x,y]$ respectively. Its Hermitian
metric is given by $\langle x|y\rangle=(x^*,y)$ where star means
complex conjugation.   In this model spin projection operator onto
direction $\ell\in\mathbb{E}^3$ is given by equation
$$J_\ell\psi=i[\ell,\psi].$$
It has real eigenstate $|0\rangle=\ell$ and two complex conjugate
ones $|\pm1\rangle=\frac{1}{\sqrt{2}}(m\pm in)$, where
$(\ell,m,n)$ is orthonormal basis of $\mathbb{E}^3$. The latter
states are {\it coherent\/}, see Example \ref{spin1}.
%In general
They may be identified with isotropic vectors
$$\psi\mbox{ is coherent}\Longleftrightarrow (\psi,\psi)=0.$$
 Their properties are
drastically different from real vectors $\ell\in\mathbb{E}^3$
called {\it completely entangled\/} spin states.
%Let me stress
%that this is not a question of mathematical definition, to be
%given later in $n^\circ$ \ref{comp_ent}. For biphton, an optical
%version of spin 1 system, such states
%%called {\it neutrally polarized\/} and
%routinely treated as entangled by experimenters \cite{Serg00}.
%(they call them {\it nonpolarized biphotons\/)}.
%General completely entangled spin sates are
%complex multiples of real vectors $\psi\in\mathbb{E}^3$.
They may be characterized mathematically as follows
%by the following property
$$\psi \mbox{ is completely entangled }\Longleftrightarrow
[\psi,\psi^*]=0.$$ Recall from Example \ref{spin1} that Lorentz
group,
%isomorphic to
being complexification of $\mathrm{SO\,}(3)$,
%acts in $\mathcal{H}$ by transformations
preserves the bilinear form $(x,y)$. Therefore
%Lorentz transformation of
it transforms a coherent state into another coherent state.
%is again coherent.
This however fails for completely entangled states.
%into coherent coherent states are Lorentz
%invariant, while completely entangled ones are not.

%{\footnotesize I have to avoid reference to Lorentz group and use
%linear optic elements instead, see $n^\circ$ \ref{gen_coh}.}
 Every
noncoherent state can be transformed into completely entangled one
by a Lorentz boost. In this respect Lorentz group
%is
plays r\^ole similar to SLOCC transform for two qubits which
allows to filter out a nonseparable state into a completely
entangled Bell state, cf. Example \ref{SLOCC}.
%acts in
%spin space $\mathcal{H}$ and transforms coherent state into
%coherent. but may change the amount of entanglement

By a rotation every spin 1 state can be put into the {\it
canonical form}
%$\bigstar$
\begin{equation}\label{canonic}\psi=m\cos\varphi+in\sin\varphi,\quad 0\le\varphi\le
\frac{\pi}{4}.
\end{equation}
The angle $\varphi$, or {\it generalized concurrence\/}
$\mu(\psi)=\cos2\varphi$,
%(\ref{concur}),
is unique intrinsic parameter of spin 1 state. The extreme values
$\varphi=0,\pi/4$ correspond to completely entangled and coherent
states respectively.

Observe that
$$J_\ell^2\psi=-[\ell,[\ell,\psi]]=\psi-(\ell,\psi)\ell$$
so that
$$S_\ell=2J_\ell^2-1:\psi\mapsto \psi-2(\ell,\psi)\ell$$
is reflection in plane orthogonal to $\ell$. Hence $S_\ell^2=1$
and operators $S_\ell$ and $S_m$ commute iff $\ell\perp m$.

Consider now a cyclic quintuplet of unit vectors
$\ell_i\in\mathbb{E}^3$, $i
\mathrm{\;mod\;} 5$,
 such that $\ell_i\perp\ell_{i+1}$, and call it {\it pentagram\/}. Put $S_i:=S_{\ell_i}$. Then
$[S_i,S_{i+1}]=0$ and for all possible values $s_i=\pm1$ of
observable  $S_i$ the following inequality holds
\begin{equation}\label{spin_bell}
s_1s_2+s_2s_3+s_3s_4+s_4s_5+s_5s_1+3\ge0.
\end{equation} Indeed product of the
monomials $s_is_{i+1}$ is equal to $+1$, hence at least one of
them is $+1$, and the sum of the rest is $\ge-4$.

%By commutativity $[S_i,S_{i+1}]=0$
Being commutative, observables $S_i,S_{i+1}$ have a joint
distribution. If all $S_i$ would have a hidden joint distribution
then taking average of (\ref{spin_bell}) one get Bell's type
inequality
\begin{equation}\label{pent_bell}
\langle\psi|S_1S_2|\psi\rangle+\langle\psi|S_2S_3|\psi\rangle+
\langle\psi|S_3S_4|\psi\rangle+
\langle\psi|S_4S_5|\psi\rangle+
\langle\psi|S_5S_1|\psi\rangle+3\ge0
\end{equation}
for testing classical realism. Note that all marginal constraints
can be obtained from
%(\ref{spin_bell})
this inequality by flips $S_i\mapsto\pm S_i$. Using equation
$S_i=1-2|\ell_i\rangle\langle\ell_i|$ one can recast it into
geometrical form
\begin{equation}\label{spin_bell1}
\sum_{i\mathrm{\;mod\;}5}|\langle\ell_i,\psi\rangle|^2\le 2\Longleftrightarrow\sum_{i\mathrm{\;mod\;} 5}\cos^2\alpha_i\le 2,\qquad\alpha_i=\widehat{\ell_i\psi}.
\end{equation}
%This inequality can be easily violated for entangled (= real)
%state $\psi$.
Completely entangled spin states easily violate this inequality.
Say for regular pentagram and $\psi\in\mathbb{E}^3$ directed along
its axis of symmetry one gets
$$\sum_{i\mathrm{\;mod\;}5}\cos^2\alpha_i=\frac{5\cos\pi/5}{1+\cos\pi/5}\approx2.236>2.$$
We'll see below that in a smaller extend every non-coherent spin
state violates inequality (\ref{spin_bell1}) for an appropriate
pentagram. The coherent states, on the contrary, pass this test
for any pentagram.
%In other words all non-coherent spin sates are
%nonclassical. They are
%%just
%{\it entangled\/} spin states
%%as defined in
%in terminology of $n^\circ$ \ref{gen_ent}.

To prove these claims write inequality (\ref{spin_bell1}) in the
form
%consider sum of projectors onto directions $\ell_i$
$$\langle\psi|A|\psi\rangle\le2,\qquad A=\sum_{i\mathrm{\;mod\;}5}|\ell_i\rangle\langle\ell_i|,$$
and observe the following properties of spectrum
$\lambda_1\ge\lambda_2\ge\lambda_3\ge0$ of operator $A$.
%\begin{claim}
\begin{enumerate}
\item $\mathrm{Tr\,} A=\lambda_1+\lambda_2+\lambda_3=5.$
\item If the pentagram contains parallel vectors $\ell_i\|\ell_j$
then $\lambda_1 =\lambda_2=2, \lambda_3=1$.
%We'll call such configuration {\it degenerate}.
\item For any pentagram with no parallel vectors
%with equality iff $\ell_i\|\ell_j$ for some $i\ne j$.
\begin{enumerate}
\item \quad $\lambda_1> 2$,
\item \quad $\lambda_3>1$,
%with equality iff $\ell_i\|\ell_j$ for some $i\ne j$.
\item \quad$\lambda_2<2$.
\end{enumerate}
\end{enumerate}
%\end{claim}
\proof (1) $\mathrm{Tr\,} A=\sum_{i\mathrm{\;mod\;} 5}\mathrm{Tr\,} |\ell_i\rangle\langle\ell_i|=5$.\\[1mm]
(2) Let say $\ell_1=\pm\ell_3$ then $\ell_3,\ell_4,\ell_5$ form
orthonormal basis of $\mathbb{E}^3$. Hence $A$ is sum of identical
operator
$|\ell_3\rangle\langle\ell_3|+|\ell_4\rangle\langle\ell_4|+|\ell_5\rangle\langle\ell_5|$
and projector
$|\ell_1\rangle\langle\ell_1|+|\ell_2\rangle\langle\ell_2|$ onto
plane
$<\ell_1,\ell_2>$.\\[1mm]
(3a) Take unit vector
$$x\in<\ell_1,\ell_2>\cap<\ell_3,\ell_4>$$ so that
$x=\langle\ell_1,x\rangle\ell_1+\langle\ell_2,x\rangle\ell_2=
\langle\ell_3,x\rangle\ell_3+\langle\ell_4,x\rangle\ell_4$.
Then
$$Ax=\langle\ell_1,x\rangle\ell_1+\langle\ell_2,x\rangle\ell_2+\langle\ell_3,x\rangle\ell_3+\langle\ell_4,x\rangle\ell_4
+\langle\ell_5,x\rangle\ell_5=2x+\langle\ell_5,x\rangle\ell_5$$
and $\lambda_1\ge\langle x|A|x\rangle=2+|\langle
x|\ell_5\rangle|^2> 2$.\\[1mm]
(3b) This property is more subtle. It amounts to positivity of the
form
$$B(x,y)=\langle x|A-1|y\rangle=\sum_{i\mathrm{\;mod\;}5}\langle x|\ell_i\rangle\langle
\ell_i|y\rangle-\langle x|y\rangle.$$ One can show that
$$\det B=2\det A\prod_{i<j}\sin^2(\widehat{\ell_i\ell_j}).$$
This implies that $B$ is nondegenerate for every pentagram of
noncollinear vectors. Therefor $B$ has the same inertia index for
all such pentagrams. Finally one can check that for
regular pentagram $B$ is positive.\\[1mm]
(3c) Follows from (1), (3a), and (3b). \qed
\begin{thm} Bell's inequality $\langle\psi|A|\psi\rangle\le2$ holds for coherent state $\psi$
and any pentagram, while non-coherent state violates this
inequality for some pentagram.
\end{thm}
\proof Take
$\psi=m\cos\varphi+in\sin\varphi$, $0\le\varphi\le\pi/4$ in
canonical form (\ref{canonic}). Then
$$\langle\psi|A|\psi\rangle=\langle m|A|m\rangle \cos^2\varphi+\langle n|A|n\rangle
\sin^2\varphi.$$
To violate Bell's inequality we have to make the right hand side
maximal. This happens for $m=|\lambda_1\rangle$, the eigenvector
of $A$ with maximal eigenvalue $\lambda_1$, and
$n=|\lambda_2\rangle$. The maximal value thus obtained is
\begin{equation}\label{max}\langle\psi|A|\psi\rangle_{\max}=\lambda_1\cos^2\varphi+\lambda_2\sin^2\varphi=\frac{\lambda_1+\lambda_2}{2}+
\frac{\lambda_1-\lambda_2}{2}\cos2\varphi.
\end{equation}
For coherent state $2\varphi=\pi/2$ we arrive at Bell's inequality
$$\langle\psi|A|\psi\rangle_{\max}=\frac{\lambda_1+\lambda_2}{2}=
\frac{5-\lambda_3}{2}\le2$$
which holds for all pentagrams by property (3b). The other part of
the theorem follows from the following
\begin{claim*} For every noncoherent state $0\le\varphi<\pi/4$
there exists pentagram s.t.
$$\langle\psi|A|\psi\rangle_{\max}=\frac{\lambda_1+\lambda_2}{2}+
\frac{\lambda_1-\lambda_2}{2}\cos2\varphi>2.$$
\end{claim*}
Indeed, for degenerate pentagram $\Pi$, containing parallel
vectors, the corresponding operator $A$ has multiple eigenvalue
$\lambda_1=\lambda_2=2$ and simple one $\lambda_3=1$. In this case
equation (\ref{max}) amounts to
%Bell's inequality degenerates into equaton
$\langle\psi|A|\psi\rangle_{\max}=2$. Let $\widetilde{A}$
be operator corresponding to a small nondegenerate
$\varepsilon$-perturbation $\widetilde{\Pi}$ of pentagram ${\Pi}$,
and $\widetilde{\lambda}$ be its spectrum. Then for simple
eigenvalue $\lambda_3$ we have by property (3b)
$$\Delta(\lambda_3)=\widetilde{\lambda}_3-\lambda_3=O(\varepsilon)>0,$$
and hence
$$\Delta(\lambda_1+\lambda_2)=\Delta(5-\lambda_3)=O(\varepsilon)<0.$$
Hereafter $O(\varepsilon)$ denote a quantity of {\it exact\,}
order
 $\varepsilon$. The increment of multiple roots $\lambda_1,\lambda_2$ is
of smaller order
$$\Delta(\lambda_1)=O(\sqrt{\varepsilon})>0,\quad\Delta(\lambda_2)=O(\sqrt{\varepsilon})<0,
\quad\Delta(\lambda_1-\lambda_2)=O(\sqrt{\varepsilon})>0,$$ where the signs of the
increments are
%determined
derived from properties (3a) and (3c). As result
$$\Delta(\langle\psi|A|\psi\rangle_{\max})=\Delta\left(\frac{\lambda_1+\lambda_2}{2}+
\frac{\lambda_1-\lambda_2}{2}\cos2\varphi\right)=
O(\varepsilon)+O(\sqrt{\varepsilon})=O(\sqrt{\varepsilon})>0,$$
provided $\cos2\varphi>0$ and $\varepsilon\ll 1$. Hence for
noncoherent state Bell's inequality fails:
$\langle\psi|\widetilde{A}|\psi\rangle_{\max}>2$.
\qed
\begin{rem}\label{squares} Product of orthogonal reflections
$S_iS_{i+1}$ in pentagram inequality (\ref{pent_bell}) is a
rotation by angle $\pi$ in plane $<\ell_i,\ell_{i+1}>$, i.e.
$$S_iS_{i+1}=1-2J^2_{[\ell_i,\ell_{i+1}]},$$
and the inequality can be written in  the form
$$
\langle\psi|J^2_{[\ell_1,\ell_{2}]}|\psi\rangle+
\langle\psi|J^2_{[\ell_2,\ell_{3}]}|\psi\rangle+
\langle\psi|J^2_{[\ell_3,\ell_{4}]}|\psi\rangle+
\langle\psi|J^2_{[\ell_4,\ell_{5}]}|\psi\rangle+
\langle\psi|J^2_{[\ell_5,\ell_{1}]}|\psi\rangle\le4.
$$
Observe that $\ell_i,\ell_{i+1},[\ell_1,\ell_{i+1}]$ are
orthogonal and therefor
$$J^2_{\ell_i}+J^2_{\ell_{i+1}}+J^2_{[\ell_i,\ell_{i+1}]}=2.$$
This allows return to operators $J_i=J_{\ell_i}$
$$
\langle\psi|J^2_1|\psi\rangle+
\langle\psi|J^2_2|\psi\rangle+
\langle\psi|J^2_3|\psi\rangle+
\langle\psi|J^2_4|\psi\rangle+
\langle\psi|J^2_5|\psi\rangle\ge3.
$$
The last inequality can be tested  experimentally by {\it
measuring\/} $J$ and {\it calculating\/} the average of $J^2$.
Thus we managed to test classical realism in framework of spin 1
dynamical system in which no two operators $J\in \frak{su\,}(2)$
commutes, cf. Example \ref{class_real}. The trick is that {\it
squares\/} of the operators may commute.

%This return us back in framework of spin 1 dynamical system. Note
%that we managed too test classical realism by measuring spin
%projections no two of which commute,
%in spite of the fact that
%algebra $su(2)$ contains no nonproportional commuting operators.
\end{rem}
\begin{rem}\label{non_class} The difference between coherent and
entangled spin states disappears for the full group
$\mathrm{SU\,}(\mathcal{H})$. Hence with respect to this group all
states are nonclassical, provided $\dim\mathcal{H}\ge 3$, cf.
\cite{Peres}.
\end{rem}

\subsection{Call for new approach}
Putting aside highly publicized philosophical aspects of
entanglement, its physical manifestation usually associated with
two phenomena:
\begin{itemize}
\item {\it violation of classical realism,}
\item {\it nonlocality.}
\end{itemize}
As we have seen above every state of a system of dimension $\ge3$
with full dynamical group $\mathrm{SU\,}(\mathcal{H})$ is
nonclassical. Therefor violation of classical realism is a general
feature of quantum mechanics in no way specific for entanglement.

The nonlocality, understood as a correlation beyond  light cones
of the systems, is a more subtle and
%still remains an
enigmatic effect. It tacitely presumes
%suppose
%that the system consists of
spatially separated components in the system. This premise
eventually ended up with formal identification of entangled states
with nonseparable ones.
%inspired by discoverers of this phenomenon
%\cite{EPR, Sch},
%its rigorous treatment based on two physical
%effects: nonlocality and violation of classical realism.
% Original approach to the subject emerges from
%detailed study of a simplest two-qubits system
%%falls under heavy influence of a
%\cite{EPR}, for which all known rigorous definitions of entangled
%state
%\marginpar{\footnotesize Rewrite this more accurately.}
%\begin{enumerate}
%\item {\it nonlocality\/} including violation of classical realism
%\cite{Bell},
%\item {\it nonclassical EPR correlation\/} \cite{Wer},
%\item {\it nonseparability\/},
%\end{enumerate}
%or their variations.
%Neither of them, however
The whole understanding  of entanglement
%falls
was formed under heavy influence
%of very degenerate case
of two-qubits, or more generally two-components systems, for which
{\it Schmidt decomposition\/} (\ref{Schmidt}) gives a transparent
description and quantification %(\ref{entropy})
of entanglement.
% and provides a natural measure for entanglement \cite{BBPS}.
%For such systems the above definitions are in strict conformity.
%where the above definitions are in strict conformity and there is
%a {\it natural} measure of entanglement \cite{BBPS}.
However later on it became clear that entanglement does manifest
itself in systems with no clearly separated components, e.g.
%where none of the above approaches make sense, including
\begin{itemize}
\item Entanglement in an ensemble of identical bosons or fermions \cite{Legg, GM04a, GM04b,
SCKLL01, ESBL02, LNP05, VBKOS04, VC03, Z02, ON02}.
\item Single particle entanglement, or entanglement of internal
degrees of freedom, see \cite{CKSh05,Kim} and references therein.
\end{itemize}

%Connection of Bell's inequalities with entanglement is very
%illusive. They can be violated even for {\it coherent states\/}
%\cite {Kly}, which are opposite  to entangled ones.

Nonlocality is meaningless for a condensate of identical bosons or
fermions with strongly overlapping  wave functions. Nevertheless
we still can distinguish
%{\it separable\/}
{\it coherent\/} Bose-Einstein condensate of bosons $\Psi=\psi^N$
or Slater determinant for fermions
$\Psi=\psi_1\wedge\psi_2\wedge\ldots\wedge\psi_N$ from generic
entangled states in these systems.
%with all particles in one and
%the same state $\psi$, from generic {\it entangled\/} state $\Psi$
%= symmetric function of states of the components.
Recall, that entangled states of biphoton where extensively
studied experimentally \cite{Serg00,Serg03}, and Bell inequalities
can be violated in such simple
%one component
system as spin 1 particle, see $n^\circ$ \ref{pent}.
%This implies that
Thus non-locality, being indisputably the most striking
manifestation of entanglement, is {\it not\/} its indispensable
constituent. See also \cite{Malley04,Malley05}.
%The current understanding
%situation
%becomes even more tricky for fermions

Lack of common ground already  led to a controversy in
understanding of entanglement in bosonic systems, see $n^\circ$
\ref{coh_v_unst},
%\cite{GM04b},
and Zen question about single particle entanglement calls for a
completely novel approach.
%falls
%far
%beyond the reach of the conventional approach.

Note finally that there is no place for entanglement in von
Neumann picture, where full dynamical group $SU(\mathcal{H})$
%ensures that
makes all states equivalent, see $n^\circ$ \ref{dynamic}.
Entanglement is an effect caused by {\it superselection rules\/}
or {\it symmetry breaking\/} which reduce the dynamical group to a
subgroup $G\subset
\mathrm{SU\,}(\mathcal{H})$ small enough to create intrinsical  difference between states.
For example, entanglement in two component system
$\mathcal{H}_A\otimes\mathcal{H}_B$ comes from reduction of the
dynamical group to
$\mathrm{SU}(\mathcal{H}_A)\times\mathrm{SU}(\mathcal{H}_B)\subset
\mathrm{SU}(\mathcal{H}_A\otimes\mathcal{H}_B)$.
Therefor entanglement
%essentially relies on the dynamical group and
{\it must\/} be studied in the framework of
%a given
quantum dynamical systems.
%$G:\mathcal{H}$.
%Therefore it should be discussed in
%framework of a given quantum dynamical system $G:\mathcal{H}$.
%must be included in the
%initial set up of any discussion of entanglement.
% discussion of entanglement quantum
%dynamical system $G:\mathcal{H}$ should a starting point of any
%discussion of entanglement.
% the initial set
%up for understanding of entangleemnt should  for understanding
%entanglement must
%include a dynamical system $G:\mathcal{H}$ with dynamical group
%$G$ small enough to ensure that not all states are equivalent.
%we need
%calls for a completely novel approach.
%for a single particle entanglement.
%\subsection{dynamical symmetry approach}
\subsection{Completely entangled states}\label{comp_ent} Roughly speaking, we consider
entanglement as a manifestation of {\it quantum fluctuations\/}
%of a physical system
in a state where they come to their extreme. Specifically, we look
for states with maximal total variance
%In the previous section  we have seen  how illusive may be
%connection between ``classical realism'' and entanglement.
%Instead of ambiguous relation with ``classical realism'' we put
%forward {\it extremal property\/} of completely entangled state,
%which can be checked in all known instances, namely the maximality
%of its total variance:
$${\mathbb D}(\psi)=\sum_i
\langle\psi|X_i^2|\psi\rangle-\langle\psi|X_i|\psi\rangle^2=\max.$$
It follows from equation (\ref{c_var})
%$$\mathbb D(\psi)=\langle\lambda,\lambda+2\rho\rangle-
%\sum_i\langle\psi|X_i|\psi\rangle^2$$
that the maximum is attained for state $\psi$ with zero
expectation of all essential observables
\begin{equation}\label{ent_eqn}\framebox{\parbox{4cm}{\begin{center}
$\langle\psi|X|\psi\rangle=0,\quad \forall X\in {\frak
L}$\end{center}}}
\begin{array}{ll}\qquad\parbox{2cm}{\begin{center}\small Entanglement equation\end{center}}
%&\textrm{\scriptsize  Entanglement }\\
%&\textrm{\scriptsize equation}
\end{array}
\end{equation}
We use this condition as the definition of {\it completely
entangled\/} state and refer to it  as {\it entanglement
equation}.
Let's outline its distinctive features.\\[1mm]
%\begin{dfn}\label{Ent}
%State $\psi\in \mathcal{H}$ is said  to be {\it completely
%entangled} if all observables  $X\in
%\frak L$ have zero expectation in state $\psi$
%$$\framebox{\parbox{4cm}{\begin{center}
%$\langle\psi|X|\psi\rangle=0,\quad \forall X\in \frak
%L$\end{center}}}$$
%\end{dfn}
%\noindent{ \begin{itemize}
%\item%In other words,
$\bullet$ Equation (\ref{ent_eqn}) tells that in completely
entangled state the system is at the center of its quantum
fluctuations.
%\item
\\[1mm]
$\bullet$ This ensure maximality of the total variance, i.e.
overall level of quantum fluctuations  in the system.
%Being
%opposite in this respect to coherent states completely entangled
%states may be considered as {\it extremely nonclassical\/}.
%Therefor one can expect that they should manifest purely quantum
%effects in prominent way.
In this respect completely entangled states are opposite to
coherent ones, and may be %called
treated as {\it extremely nonclassical\/}.
%Therefor
%for such states
They should manifest purely quantum effects, like violation of
classical realism, to the utmost.
%are most likely to
%in the most prominent way.
%should manifest themselves in the most prominent way.
% in completely entangled states.
%\item
\\[1mm]
$\bullet$ May be the main
%shortcoming
flaw of the conventional approach is lack of {\it physical
quantity\/} associated with entanglement. In contrast to this, we
consider entanglement as a manifestation of {\it quantum
fluctuations\/} in a state where they come to their extreme. This,
for example, may help
%, for example,
to understand stabilizing effect of environment on an entangled
state, see \cite{CKS}.
%, if quantum fluctuations {\it decrease\/} its energy.
%Classical driving field is a good example
%This mechanism was implemented in
%\cite{CKS}.
%for entangled atoms in a cavity.
\\[1mm]
$\bullet$ Entanglement equation (\ref{ent_eqn}) and the maximality
of the total fluctuations plays an important heuristic r\^ole,
similar to variational principles in mechanics. It has also
%an important geometrical interpretation
a transparent geometrical meaning discussed below in $n^\circ$
\ref{gen_ent}. This interpretation puts entanglement in framework
of Geometric Invariant Theory, which provides powerful methods for
solving quantum informational problems \cite{Kl04}.
%\item
\\[1mm]
$\bullet$ The total level of quantum fluctuations in irreducible
system $G:\mathcal{H}_\lambda$ varies in the range
\begin{equation}\label{range} \langle\lambda,2\delta\rangle\le\mathbb
D(\psi)\le\langle\lambda,
\lambda+2\delta\rangle
\end{equation}
with minimum attained at {\it coherent\/} states, and the maximum
for {\it completely entangled\/} ones, see $n^\circ$
\ref{extr_coh}. For spin $s$ system this
%gives
amounts to  $s\le\mathbb{D}(\psi)\le s(s+1)$.
%\item Our approach suggests that {\it entanglement is a manifestation of
%quantum fluctuations in a state where they come to their extreme.}
\\[1mm]
$\bullet$ Extremely high level of quantum fluctuations makes every
completely entangled state manifestly nonclassical, see Example
\ref{nonclass} below.
\\[1mm]
$\bullet$ The above definition make sense for any quantum system
$G:\mathcal{H}$ and it is in conformity with conventional one when
the latter is applicable, e.g. for multi-component systems, see
Example \ref{multy_comp}. For spin 1 system completely entangled
spin states coincide with so called {\it neutrally polarized\/}
states of biphoton, see $n^\circ$ \ref{pent} and
\cite{Serg00,Serg03}.
%\item
\\[1mm]
$\bullet$ As expected, the definition is $G$-invariant, i.e. the
dynamical group transforms completely entangled state $\psi$ into
completely entangled one $g\psi$, $g\in G$.
%\\[1mm]
%$\bullet$
%\end{itemize}}
\begin{rem}\label{unst_syst}
%\marginpar
%{\footnotesize Actually I know no examples of completely entangled
%states for unitary representations of {\it noncompact\/} groups.
%Look first at Lorentz group $\mathrm{SL\,}(2,\mathbb{C})$. May be
%more interesting examples stem from {\it gauge theory}. I think in
%this case entanglement equation in representation of gauge group
%$G$ in space of sections of a $G$ bundle should amounts to a
%meaningful geometrical object. }

There are few systems where completely entangled states fail to
exist, e.g. in quantum system $\mathcal{H}$ with
%access to all degrees of freedom
full dynamical group $G={\rm SU}(\mathcal{H})$ all states are
coherent.
%two partite system $\mathcal H_A\otimes\mathcal H_B$
%with components of different dimension.
In this case the total variance (\ref{Tot_Var}) still attains some
maximum, but it doesn't satisfy entanglement equation
(\ref{ent_eqn}). We use different terms {\it maximally\/} and {\it
completely\/} entangled states to distinguish these two
possibilities and to stress conceptual, rather then quantitative,
origin of genuine entanglement governed by equation
(\ref{ent_eqn}). In most cases these notions are equivalent,
%All
and all exceptions are actually known, see $n^\circ$ \ref{rigid}.

To emphasize  the aforementioned difference we call quantum system
$G:\mathcal{H}$ {\it stable\/} if it contains a completely
entangled state, and {\it unstable\/} otherwise.

%Note also that for infinite dimensional systems, like quantum
%oscillator $n^\circ$ \ref{q_oscil}, the total variance may be
%infinite, while entanglement equation (\ref{ent_eqn}) still makes
%sense.

%Note also that for noncompact dynamical group $G$, like
%Weyl-Heisenberg group for quantum oscillator \ref{W-H}, faithful
%unitary representations are infinite dimensional. The total
%variance often make no sense, while entanglement equation
%\ref{ent_eqn}
%but their physical meaning is still unclear.
\end{rem}
\begin{Ex}\label{spin1_var} The conventional definition of entanglement explicitly refers to a
{\it composite} system, which from our point of view is no more
reasonable for entangled states, then for coherent ones. As an
example let's consider completely entangled state
$\psi\in\mathcal{H}_s$ of  spin $s$ system. According to the
definition this means that average spin projection onto {\it
every\/} direction $\ell$ should be  zero:
$\langle\psi|J_\ell|\psi\rangle=0$. This certainly can't happens
for $s=1/2$, since in this case all states are coherent and have
definite spin projection $1/2$ onto some direction. But for $s\ge
1$ such states do exist and will be described later in $n^\circ$
%we'll describe them later in $n^\circ$
\ref{HilMum}.
%Example \ref{ent_spin}.
For example, one can take $\psi=|0\rangle$ for integral spin $s$,
and %or in general
%it is easy to check that
$$\psi=\frac1{\sqrt{2}}(|+s\rangle-|-s\rangle)$$ for any $s\ge 1$.
%Up to a rotation this are the only possibility for $s=1$ or $3/2$.
They  have extremely big fluctuations $\mathbb{D}(\psi)=s(s+1)$,
and therefor are {\it manifestly nonclassical\/}:
%Every such state is {\it manifestly nonclassical\/},
average spin projection onto every direction is zero, while the
standard deviation $\sqrt{s(s+1)}$ {\it exceeds\/} maximum of the
spin projection $s$.

\begin{Ex}\label{nonclass} This consideration can be literally transferred to an arbitrary
irreducible system $G:\mathcal{H}_\lambda$, using inequality
$\langle\lambda,\lambda\rangle<\langle\lambda,\lambda+2\delta\rangle$
instead of $s^2< s(s+1)$, to the effect that a completely
entangled state of any system is nonclassical.
\end{Ex}
%For spin $s$ system the total variance satisfies inequality
%$$s\le\mathbb{D}(\psi)\le s(s+1).$$
%For $s=1/2$ all spin states are coherent and there are no
%completely entangled states. But for $s\ge1$ such states do exist,
%and we describe them below.
\end{Ex}

\begin{Ex}\label{multy_comp} Entanglement equation
(\ref{ent_eqn}) implies that state of a multicomponent system, say
$\psi\in\mathcal{H}_{ABC}=\mathcal{H}_{A}\otimes\mathcal{H}_{B}\otimes\mathcal{H}_{C}$,
is completely entangled iff its marginals $\rho_A,\rho_B,\rho_C$
are scalar operators. This observation is in conformity with
conventional approach to entanglemnt \cite{DVC}, cf. also Example
\ref{Schmidt_ex}.
%W.~D\"ur, G.~Vidal, and J.I.~Cirac, Phys. Rev. A, {\bf 62} (2000),
%062314.}
\end{Ex}
%\begin{Ex}  For state $\psi$ of $N$ qubit
%$$N\le\mathbb{D}(\psi)\le N\left(\frac32\right).$$
%The minimum corresponds to {\it separable states\/}, and the
%maximum to {\it completely entangled} ones.
%\end{Ex}

%In this example coherent states have definite spin projection $s$
%onto some direction:
%$$\framebox{\parbox{7cm}{\begin{center}$\psi$ is coherent $\Longleftrightarrow$
%$\psi=|s\rangle$\end{center}}}$$  Group of {\it complex}
%symmetries of such state is conjugate to group of triangular
%matrices (=Borel subgroup).

\subsection{General entangled states and stability}\label{gen_ent}

From operational point of view state $\psi\in\mathcal{H}$ is {\it
entangled} iff one can filter out from $\psi$ a completely
entangled state $\psi_0$ using SLOCC operations. As we know from
Example \ref{SLOCC} in standard quantum information  settings
SLOCC group coincide with complexification $G^c$ of the dydnamic
group $G$. This leads us to the following
\begin{dfn}\label{ent_def} State $\psi\in\mathcal{H}$ of dynamical system
$G:\mathcal{H}$ is said to be {\it entangled\/} iff it can be
transformed into a completely entangled state $\psi_0=g\psi$ by
complexified group $G^c$ (possibly asymptotically $\psi_0=\lim_i
g_i\psi$ for some sequence $g_i\in G^c$).
\end{dfn}

In Geometric Invariant Theory such states $\psi$ are called  {\it
stable} (or {\it semistable} if $\psi_0$ can be reached only
asymptotically). Their intrinsic characterization is one of the
central problems both in Invariant Theory and in Quantum
Information. Relation between these two theories can be
summarized in the following table, with some entries to be
explained below.
%We summarize this in the following table with some item explained below.

\begin{center}
{\sc Dictionary}\\*
\vspace{3mm}
{\begin{tabular}{|c|c|}
\hline
\parbox{5.1cm}{\vspace{2pt}\begin{center}\bf Quantum Information\end{center}\vspace{2pt}} &
\parbox{5.1cm}{\vspace{2pt} \begin{center}\bf Invariant Theory\end{center}\vspace{2pt}}\\
\hline\hline
\parbox{5.1cm}{\vspace{2pt}\begin{center}Entangled state\end{center}\vspace{2pt}}&
\parbox{5.1cm}{\vspace{2pt}\begin{center}Semistable vector\end{center}\vspace{2pt}}\\
\hline
\parbox{5.1cm}{\vspace{2pt}\begin{center} Disentangled state \end{center}\vspace{2pt}}&
\parbox{5.1cm}{\vspace{2pt}\begin{center}Unstable vector\end{center}\vspace{2pt}}\\
\hline
\parbox{5.1cm}{\vspace{2pt}\begin{center}SLOCC transform\end{center}\vspace{2pt}}&
\parbox{5.1cm}{\vspace{2pt}\begin{center}Action of complexified group $G^c$\end{center}\vspace{2pt}}
\\
\hline
\parbox{5.1cm}{\vspace{2pt} \begin{center}Completely entangled state $\psi_0$ prepared from
$\psi$ by SLOCC\end{center} \vspace{2pt}}&
\parbox{5.1cm}
{\vspace{2pt}\begin{center} Minimal vector $\psi_0$ in complex orbit of $\psi$\end{center}\vspace{2pt}}\\
\hline
\parbox{5.1cm}{\vspace{2pt}\begin{center}State obtained  from comp\-letely entangled one by SLOCC\end{center}\vspace{2pt}}&
\parbox{5.1cm}{\vspace{2pt}\begin{center}Stable vector\end{center}\vspace{2pt}}
\\
\hline
\end{tabular}}
\vspace{3mm}
\end{center}
Completely entangled states can be
%in complex orbit $G^c\psi=\{g\psi\mid g\in G^c\}$ are
characterized by the following theorem, known as {\it Kempf--Ness
unitary trick.}
\begin{thm}[Kempf-Ness \cite{K-N}]\label{K-N} State $\psi\in\mathcal{H}$ is
completely entangled iff it has minimal length in its complex
orbit
\begin{equation}\label{min_vec}|\psi|\le |g\cdot\psi|,\quad\forall g\in G^c.
\end{equation}
Complex orbit $G^c\psi$ contains a completely entangled state iff
it is closed. In this case the completely entangled state is
unique up to action of $G$.
\end{thm}
\begin{rem} Recall that entangled state $\psi$ can be {\it asympotically\/}
transformed by SLOCC into a completely entangled one. By
Kempf-Ness theorem the question when  this can be done
 {\it effectively\/} depends on whether the complex orbit of
$\psi$ is closed or not.
%We call such states stable.
The following result gives a necessary condition for this.
%A necessary condition for this was given by
%\marginpar
%{\footnotesize Give here Dadok-Kac
%construction of completely entangled states \cite{DadKac85}.
%According to V.~Popov \cite{Pop72} for semisimple group stability
%of action is equivalent the generic stabilizer to be reductive.
%For reductive group one needs also stability of the center acting
%in $V/[G,G]$, see \cite{Pop92} p.78. }
\end{rem}
\begin{thm*}[Matsushima
\cite{Mats60}]
%For dynamical system $G:\mathcal{H}$
%$$(G^c)_\psi=(G_\psi)^c,$$
Complex stabilizer $(G^c)_\psi$ of stable state $\psi$  coincides
with complexification of its compact stabilizer $(G_\psi)^c$.
%(\ref{stabilizer}).
\end{thm*}

Square of length of the minimal vector in complex orbit
\begin{equation}\label{concur}\mu(\psi)=\inf_{g\in G^c}|g\psi|^2,
%\quad 0\leC(\psi)\le 1$,
\end{equation}
provides a natural quantification of entanglement. It amounts to
$\cos 2\varphi$ for spin 1 state (\ref{canonic}), to {\it
concurrence\/} $C(\psi)$ \cite{HW} in two qubits, and to square
root of {\it 3-tangle\/} $\tau(\psi)$ for three qubits (see
below). We call it {\it generalized concurrence\/}. Evidently
$0\le\mu(\psi)\le1$.
%\marginpar
%{\footnotesize It's not a good idea call $\mu$ concurrence,
%which should be a linear function of the variance, see
%\texttt{baris-APL.tex.}}
%It varies from $0$ to $1$ and attaines its

Equation $\mu(\psi)=1$ tells that $\psi$ is already  a minimal
vector, hence completely entangled state.
%only for completely entangled

Nonvanishing of the generalized concurrence $\mu(\psi)>0$ means
that  closure of complex orbit $\overline{G^c\psi}$ doesn't
contains zero. Then the  orbit of minimal dimension
$\mathcal{O}\subset\overline{G^c\psi}$ is closed and nonzero.
Hence by Kempf-Ness unitary trick it contains a completely
entangled state $\psi_0\in\mathcal{O}$ which asymptotically can be
obtained from $\psi$ by action of the complexified dynamical
group. Therefor by definition \ref{ent_def}
$$\mu(\psi)>0\Longleftrightarrow\mbox{ $\psi$ is entangled.}$$
\subsection{Coherent versus unstable states}\label{coh_v_unst}
The minimal value $\mu(\psi)=0$ corresponds to  {\it unstatable\/}
vectors that can asymptotically fall into zero under action of the
complexified dynamical group. They form so-called {\it null
cone\/}. It contains all coherent states, along with some others
degenerate states, like $W$-state in three qubits, see Example
\ref{class_ex}.
%The latter produce
%cause
%source of

Noncoherent unstable states cause  many controversies. There is
unanimous agrement that coherent states are disentangled.
%Coherent states are unanimously recognized as disentangled.
%There is no
%doubt that coherent states are disentangled.
In approach pursued in \cite{VBKOS04} all noncoherent states are
treated as entangled. Other researchers \cite{GM04a, GM04b} argue
that some noncoherent unstable bosonic states are actually
disentangled.
%In our operational approach all
From our operational point of view all unstable states should be
%considered
treated as disentangled, since they can't be filtered out into a
completely entangled state even asymptotically.
%A warning is in order:
Therefore we accept the equivalence
%\begin{center}
%\framebox{
$$\mbox{\sc disentangled }\Longleftrightarrow \mbox{\sc unstable } \Longleftrightarrow \mbox{\sc not {
semi\/}stable. }$$
%}\end{center}
%\begin{rem}
\subsubsection{Systems in which all unstable states are coherent}\label{coh=unst}
The above controversy
%becomes void
vanishes %away
iff the null cone
%consists of
contains only coherent states,
%Or equivalently,
or equivalently %complexified
dynamical group $G$ acts transitively
%on the null cone.
on unstable states. Spin one and two qubits systems are the most
notorious examples. They are low dimensional {\it orthogonal
systems\/} with dynamical group $\mathrm{SO\,}(n)$ acting in
$\mathcal{H}^n=\mathbb{E}^n\otimes{\mathbb{C}}$ by Euclidean
rotations. Null cone in this case consists of isotropic vectors
$(x,x)=0$, which are at the same time  coherent states, cf.
$n^\circ$ \ref{pent}.
%Among  {\it simple\/} dynamical groups there
%are in addition only two exceptional irreducible systems
%$G:\mathcal{H}$ with this property.
\begin{thm} Stable irreducible system $G:\mathcal{H}$ in which all unstable states
are coherent is one of the following
%Suppose that irreducible dynamical system %$G:\mathcal{H}$
%admits entangled states and  all its unstable states are coherent.
%Then the system  is one is one of the following
\begin{itemize}
\item Orthogonal system $\mathrm{SO\,}(\mathcal{H}):\mathcal{H}$,
\item Spinor
representation of group ${\mathrm{Spin\,}}(7)$ of dimension 8,
\item
Exceptional group $G_2$ in its fundamental representation of
dimension 7.
%\item Spinor
%representation of group ${\mathrm{Spin\,}}(9)$ of dimension 16.
\end{itemize}
\end{thm}

{\footnotesize The theorem can be deduced from Theorem \ref{Licht}
%Lichtenstein's characterization of
characterizing coherent states by quadratic equations.
%\begin{equation}\label{Licht_eqn}
%C(\psi\otimes\psi)=\langle2\lambda,2\lambda+2\delta\rangle(\psi\otimes\psi),
%\end{equation}
%where $C$ is Casimir operator.
Indeed, the null cone is given by
vanishing of all invariants. Hence
%by (\ref{Licht_eqn})
in conditions of the theorem the fundamental invariants should
have degree two. For irreducible representation there is at most
one invariant of degree two, the invariant metric $(x,y)$. Thus
the problem reduces to description of subgroups
$G\subset\mathrm{SO\,}(\mathcal{H})$ acting transitively on
isotropic cone $(x,x)=0$. The metric $(x,x)$ is unique basic
invariant of such system.
%and the cone of coherent states has codimension one.
Looking into the table in Vinberg-Popov book \cite{VP} we find
only one indecomposable system with unique basic invariant of
degree two not listed in the theorem: spinor representation of
$\mathrm{Spin\,}(9)$ of dimension 16 studied by Igusa
\cite{Igusa70}. However, as we'll see below, the action of this
group $\mathrm{Spin\,}(9)$ on the isotropic cone is not
transitive.

Coherent states of decomposable irreducible system $G_A\times
G_B:\mathcal{H}_A\otimes\mathcal{H}_B$ are products
$\psi_A\otimes\psi_B$ of coherent states of the components. Hence
codimension of the cone of coherent states is at least
$d_Ad_B-d_A-d_B+1=(d_A-1)(d_B-1)$. As we've seen above, in
conditions of the theorem the codimension should be equal to one,
which is possible only for system of two qubits $d_A=d_B=2$, which
is equivalent to orthogonal system of dimension four. One can also
argue that projective quadric $Q:(x,x)=0$ of dimension greater
then two is indecomposable $Q\neq X\times Y$.
\qed}

 Both exceptional systems carry
an invariant symmetric form $(x,y)$. Scalar square $(x,x)$
generates the algebra of invarinats, and  therefore the null cone
consists of isotropic vectors $(x,x)=0$, as in the orthogonal
case.
%They appear also in
These mysterious systems emerge  also
%in  differential geometry
as exceptional {\it holonomy groups\/} of Riemann manifolds
\cite{Ber03}. Their physical meaning is unclear.
%Another exceptional system
%\marginpar
%{\footnotesize
%% Add description of the second
%%system as restriction of the standard vector representation of
%%$\mathrm{Spin\,}(7)$ onto its intersection with a conjugate subgroup in
%%$\mathrm{Spin\,}(8)$ under Cartan's triality automorphism.
%Cone of the coherent states in any irreducible representation
%$\mathcal{H}_\lambda$ is an intersection of quadrics, see
%\cite{Licht82}. Specifically, coherent states are defined by
%equation
%$C(\psi\otimes\psi)=\langle2\lambda+2\delta,2\lambda\rangle(\psi\otimes\psi)$,
%where $C$ is the Casimir operator.}

\'Elie Cartan \cite{Cart66} carefully studied coherent
states in irreducible (half)spinor representations of
$\mathrm{Spin\,}(n)$ of dimension $2^\nu$, $\nu
=\lfloor\frac{n-1}{2}\rfloor$. He call them {\it pure spinors\/}.
In general the cone of pure spinors is intersection of
$2^{\nu-1}(2^\nu+1)-{2\nu+1\choose\nu}$ linear independent
quadrics.

For $n<7$ there are no equations, i.e. all states are coherent. In
such systems there is no entanglement whatsoever, and we exclude
them from the theorem. These systems are very special and have a
transparent physical interpretation.
\begin{itemize}
\item For $n=3$ spinor representation of dimension two identifies
$\mathrm{Spin\,}(3)$ with $\mathrm{SU\,}(2)$. Vector
representation of $\mathrm{SO\,}(3)$ is just spin 1 system,
studied in $n^\circ$ \ref{pent}.

\item Two dimnensional halfspinor representations of $\mathrm{Spin\,}(4)$ identify
this group with $\mathrm{SU\,}(2)\times\mathrm{SU\,}(2)$ and the
orthogonal system of dimension 4 with two qubits.

\item For $n=5$ spinor representation $\mathcal{H}^4$ of dimension 4 carries invariant
simplectic form $\omega$ and identify $\mathrm{Spin\,}(5)$ with
simplectic group $\mathrm{Sp\,}(\mathcal{H}^4,\omega)$. The
standard vector representation of $\mathrm{SO\,}(5)$ in this
settings can be identified with the space of skew symmetric forms
in $\mathcal{H}^4$ modulo the defining form $\omega$.
%of is given by irreducible component of
%dimention 5 in $\wedge^2\mathrm{Sp\,}(4)$.

\item For $n=6$ halfsinor representations of dimension 4 identify
$\mathrm{Spin\,}(6)$ with $\mathrm{SU\,}(\mathcal{H}^4)$ and the
orthogonal system of dimension 6 with
$\mathrm{SU\,}(\mathcal{H}^4):\wedge^2\mathcal{H}^4$. This is a
system of two fermions of rank 4. The previous group
$\mathrm{Spin\,}(5)\simeq\mathrm{Sp\,}(\mathcal{H}^4)$ is just a
stabilizer of a generic state $\omega\in\wedge^2\mathcal{H}^4$.
%\marginpar{\footnotesize In forward to Cartan's book \cite{Cart66}
%it is mentioned $O(3,3)$ and $U(4)$ theories by Salam, Delboirgo,
%Strathdee of strong interactions.}
\end{itemize}

In the next case $n=7$ coherent states are defined  by single
equation $(x,x)=0$ and coincide with unstable ones. Thus we arrive
at the
%Along with the
first special  system $\mathrm{Spin\,}(7):\mathcal{H}^8$.

Stabilizer of a non isotropic spinor $\psi\in\mathcal{H}^8$,
$(\psi,\psi)\ne 0$ in $\mathrm{Spin\,}(7)$ is exceptional group
$G_2$ and its representation in orthogonal complement to $\psi$
gives the second system $G_2:\mathcal{H}^7$. Alternatively it can
be described as representation of automorphism group of Cayley
octonic algebra in the space of purely imaginary octaves.

Halfspinor representations of $\mathrm{Spin\,}(8):\mathcal{H}^8$
also carry invariant symmetric form $(x,y)$.
%and coherent states are given by equation $(x,x)=0$.
It follows  that $\mathrm{Spin\,}(8)$ acts on halfspinors as {\it
full\/} group of
%the spinor group acts as group of {\it all}
orthogonal transformations. Hence these representations are
geometrically equivalent to the orthogonal system
$\mathrm{SO\,}(\mathcal{H}^8):\mathcal{H}^8$. The equivalence is
known as Cartan's {\it triality\/} \cite{Cart66}.

Finally spinor representation of $\mathrm{Spin\,}(9)$ of dimension
16 also carries invariant symmetric form $(x,y)$ which is unique
basic invariant of this representation. However according to
Cartan's formula the cone of pure spinors is intersection of 10
independent quadrics, hence differs from the null cone $(x,x)=0$.

%Coherent states coincide with isotropic vectors $(x,x)=0$ also for
%halfspinor representations of $\mathrm{Spin\,}(8):\mathcal{H}^8$. In this
%case group dimension 8 halfspinors can be transformed into vectors
%by Cartan's trialty automorphism \cite{Cart66}. and therefor  are
%geometrically equivalent to the orthogonal system
%$\mathrm{SO\,}(8):\mathcal{H}^8$.
%
%%this gives
%together with two
%%eight dimensional
%halfspinor representations of $\mathrm{Spin\,}(8)$, which are geometrically
%equivalent to the standard vector representation of $\mathrm{SO\,}(8)$
%%standard representation of $\mathrm{SO\,}(8)$
%due to Catran's {\it triality\/}, see \cite{Cart66}. Note that
%group $\mathrm{SO\,}(8)$ contains {\it three\/} subgroups isomorphic to
%$\mathrm{SO\,}(7)$ and conjugate by the triality automorphism. Any two of
%them intersect along subgroup $G_2\subset\mathrm{SO\,}(7)$ which gives the
%second exceptional system. Another interpretation: The exceptional
%group $G_2$ is  stabilizer of a generic spinor $\psi$ in
%$\mathrm{Spin\,}(7)$ and its representation in orthogonal complement to
%$\psi$ gives the second exceptional system.
%%The second one
%Alternatively it can be described as representation of
%automorphism group of Cayley octonic algebra in space of purely
%imaginary octaves.
%by an exterior automorphism.
\subsubsection{Fermionic realization of spinor representations}
%\marginpar
%{\footnotesize Actually one have to take {\it skew
%symmetrized\/} products for fermions and {\it symmetrized\/} for
%bosons to avoid scalars. Tell about algebra  $\frak{sp\,}(2n)$ for
%bosons which is responsible for polarization of $n$-photons (?).
%It can't be lifted to representation of group, see Exercise 14.2
%in \cite{Hurt83}}

Spinor representations of two fold covering $\mathrm{Spin\,}(2n)$
of orthogonal group $\mathrm{SO\,}(2n)$ have a natural {\it
physical realization\/}. Recall that all {\it quadratic\,}
expressions in creation and annihilation operators $a_i^\dag,a_j$,
$i,j=1\ldots n$ in a system of fermions with $n$ intrinsic degrees
of freedom form orthogonal Lie algebras $\frak{so\,}(2n)$
augmented by scalar operator (to
%get rid of
avoid scalars one have to use $\frac{1}{2}(a_i^\dag
a_i-a_ia_i^\dag)$ instead of $a_i^\dag a_i$, $a_ia_i^\dag$).
%Recall that all
%commutators of creation and annihilation operators
%$$[a_i^\dag,a_j^\dag],\quad [a_i^\dag,a_j],\quad [a_i,a_j],\qquad i,j=1\ldots n$$
%{\it quadratic\,}
%combinations of
%creation and annihilation operators $a_i^\dag, a_j$
%for a system of  fermions with intrinsic  state space of dimension
%$n$ form orthogonal Lie algebra $\frak{so\,}(2n)$.
%acting in the fermionic Fock
%space, known also as Clifford algebra $C(2n)$.
It acts in fermionic Fock space $\mathbb{F}(n)$, known as {\it
spinor representation\,} of $\frak{so\,}(2n)$. In difference with
bosonic case
%this representation
it has finite dimension $\dim\mathbb{F}(n)=2^n$ and splits into
two {\it halfspinor\,} irreducible components
$\mathbb{F}(n)=\mathbb{F}_{ev}(n)\oplus\mathbb{F}_{odd}(n)$,
containing even and odd number of fermions respectively.
%and
%called {\it halfspinor\,} representations.

%\marginpar{\footnotesize
%Fermionic Fock space is actually irreducible representations of
%Clifford algebra $C(2n)$  generated by determinant $\wedge^nV\in
%C(2n)$ of maximal isotropic subspace $V$, and splits into two
%halfspinor representations consisting of even and odd elements
%when restricted on $C_{ev}(2n)$.}
%It is the {\it spinor representation\,} of $\frak{so\/}(2n)$.
%It splits into two irreducible componets $C_{ev}(2n)\oplus
%C_{odd}(2n)$, containing even and odd number of fermions
%respectively, and called {\it halfspinor\/} representations.
For fermions of dimension $n=4$ the halfspinors can be transformed
into vectors
%differ from vectors
%these representations differ from the standard one
by the Cartan's triality. This provides a physical interpretation
of the orthogonal system of dimension 8.

%As another physical example of orthogonal system consider two
%fermions of dimension 4. Its state space
%$\wedge^2\mathcal{H}^4=\mathcal{H}^6$ carries invariant  quadratic
%form, given by wedge product. This identifies two fermion system
%$\mathrm{SU\,}(\mathcal{H}^4):\wedge^2\mathcal{H}^4$ with orthogonal one
%$\mathrm{SO\,}(\mathbb{E}^6):\mathcal{H}^6$.
To sum up, orthogonal systems of dimension $n=3,4,6,8$ have the
following physical description
%interpretation
\begin{itemize}
\item $n=3.$\quad Spin 1 system.
\item $n=4.$\quad Two qubit system.
\item $n=6.$\quad System of two fermions $\mathrm{SU\,}(\mathcal{H}^4):\wedge^2\mathcal{H}^4$
of dimension $4$.
\item $n=8.$\quad System of fermions  of dimension 4 with variable number of
particles (either even or odd).
\end{itemize}
%\marginpar{\tiny What
%may be classical analogue of coherent state in the fermionic
%system as $n\rightarrow\infty$? An interesting example
%%of such system may be
%is Dirac interpretation of anti particles as holes in the see
%fermions occupying states with negative energy. Note that
%restriction of halfspinors $\mathcal{H}_\pm(n+m)$ of
%$\mathrm{Spin\,}(2n+2m)$ onto
%$\mathrm{Spin\,}(2n)\times\mathrm{Spin\,}(2m)$ splits as follows:
%$\mathcal{H}_+(n+m)=\mathcal{H}_+(n)\otimes\mathcal{H}_+(m)+\mathcal{H}_-(n)\otimes\mathcal{H}_-(m)$,
%$\mathcal{H}_-(n+m)=\mathcal{H}_+(n)\otimes\mathcal{H}_-(m)+\mathcal{H}_-(n)\otimes\mathcal{H}_+(m)$.
%See \cite{Simon97} and $n^\circ$ \ref{gen_coh} for more on
%fermionic coherent states.}
The last example is fermionic analogue
of a system of quantum oscillators $n^\circ$ \ref{q_oscil}.
%We'll study entanglement in
%orthogonal systems in  section \ref{orth_sys}.
Lack of the aforementioned controversy makes description of pure
and mixed entanglement in orthogonal systems very transparent, and
quite  similar to that of two qubit and spin 1 systems, see
$n^\circ$\ref{orth_sys}.

%The latter orthogonal system has
\subsection{Unstable systems}\label{rigid}
%Here we consider another extremal
%%systems $G:\mathcal{H}$
%where all states are unstable, i.e. there is no genuine entangled
%states satisfying equation (\ref{ent_eqn}).
Halfspinor representations of the next group $\mathrm{Spin\,}(10)$
was discussed as an intriguing possibility, that quarks and
leptons may be composed of five different species of fundamental
fermionic objects \cite{Zee, WilZee82}. This is a very special
system where all states are unstable, hence
%there is
disentangled. In other words the null cone amounts to the whole
state space and  there is no genuine entanglement  governed by
equation (\ref{ent_eqn}). Such systems are opposite to those
considered in the preceding section, where the null cone is as
small as possible. We call them {\it unstable\/}. There are very
few types of such indecomposable irreducible dynamical systems
\cite{VP, Mum}:
\begin{itemize}\item %Standard representations
Unitary  system $\mathrm{SU\,}(\mathcal{H}):\mathcal{H}$;
\item Symplectic system
$\mathrm{Sp\,}(\mathcal{H}):\mathcal{H}$;
\item System of two fermions $\mathrm{SU\,}(\mathcal{H}):\wedge^2\mathcal{H}$
of odd dimension
%odd number   of internal degrees of freedom
$\dim\mathcal{H}=2k+1$;
%in space of skew-symmetric bilinear forms.
\item A halfspinor representation of dimension 16 of
$\mathrm{Spin\,}(10)$.
\end{itemize}
All (half)spinor irreducible representations for $n<7$ fall into
this category. There are many more such composite systems, and
their classification is also known due to M.~Sato and T.~Kimura
\cite{SK}.

%Physical meaning of the two exceptional systems is unclear. They
%appear also in differential geometry as exceptional holonomy
%groups of Riemann manifolds \cite{Ber03}.
%%Another exceptional system
%The second one can be described mathematically as automorphism
%group of Cayley algebra acting in space of purely imaginary
%octaves.
%Its physical meaning is unclear.

%\subsection{Orthogonal systems}\label{orth_sys}
% Here
%we consider entanglement in orthogonal systems, which
%%These systems
%may give an answer to Vollbrecht and Werner question ``Why two
%qubits are special'' \cite{Wer99}. Add  mixed entangled states in
%such systems (mixed entanglement is the essence of the Werner
%question). Try to extend results on mixed states in \cite{MKB04}
%to general systems. Include classification of nilpotent orbits of
%$\mathrm{SO\,}(\mathcal{H})\times \mathrm{SO\,}(\mathcal{H})$ in
%$\mathcal{H}\otimes\mathcal{H}$, see pp. 112+ and
%\cite{ColMcGov93}. Use it to correct
% Verstraete et al. classification of 4 qubit states \cite{Verst02}.
%\end{rem}
%\begin{Ex} For two qubits every state is either coherent
%(=separable) or entangled (=semistable). However
%\end{Ex}

\subsection{Classical criterion of entanglement}
Kempf--Ness theorem \ref{K-N} identifies closed orbits of
complexified group $G^c$ with completely entangled states modulo
action of $G$. Closed orbits can be separated by $G$-invariant
polynomials. This leads to the following {\it classical
criterion\/} of entanglement.
\begin{thm}[Classical Criterion]\label{classic}
State $\psi\in\mathcal{H}$ is entangled iff it can be separated
from zero by $G$-invariant polynomial
$$f(\psi)\neq f(0),\quad f(gx)=f(x), \forall g\in G,x\in\mathcal{H}.$$
\end{thm}
\begin{Ex} \label{class_ex}
For
%state $\psi\in\mathcal{H}_A\otimes\mathcal{H}_B$ of
two component system $\psi\in\mathcal{H}_A\otimes\mathcal{H}_B$
all invariants are polynomials in $\det[\psi_{ij}]$ (no invariants
for $\dim\mathcal{H}_A\ne\dim\mathcal{H}_B$). Hence state is
entangled iff $\det[\psi_{ij}]\ne 0$. The generalized concurrence
(\ref{concur}) related to this basic invariant by equation
$$\mu(\psi)=n|\det[\psi_{ij}]|^{2/n}.$$
%related to concurrence by equation
%$\mu(\psi)=2\left|\det[\psi_{ij}]\right|$.
Unique basic invariant for 3-qubit is {\it Cayley
hyperdeterminant} \cite{Miy,GKZ94}
\begin{eqnarray*}{\rm Det\,}[\psi]
&=&(\psi_{000}^2\psi_{111}^2+\psi_{001}^2\psi_{110}^2+\psi_{010}^2\psi_{101}^2+\psi_{011}^2\psi_{100}^2)\\
&\;&-2(\psi_{000}\psi_{001}\psi_{110}\psi_{111}+\psi_{000}\psi_{010}\psi_{101}\psi_{111}
\\&\;&+\psi_{000}\psi_{011}\psi_{100}\psi_{111}+\psi_{001}\psi_{010}\psi_{101}\psi_{110}
\\&\;&+\psi_{001}\psi_{011}\psi_{110}\psi_{100}+\psi_{010}\psi_{011}\psi_{101}\psi_{100})\\
&\;&+4(\psi_{000}\psi_{011}\psi_{101}\psi_{110}+\psi_{001}\psi_{010}\psi_{100}\psi_{111}).
\end{eqnarray*}
related to 3-tangle \cite{CKW} and generalized concurrence
(\ref{concur}) by equations
$$\tau(\psi)=4|\mathrm{Det}[\psi]|,\qquad\mu(\psi)=\sqrt{\tau(\psi)}.$$
One can check that Cayley hyperdeterminant vanishes for so called
{\it W-state\/}
$$W=\frac{|100\rangle+|010\rangle+|001\rangle}{\sqrt{3}}$$
which therefor is neither entangled nor coherent.
\end{Ex}
\begin{rem} This examples
elucidate the  nature of entanglement introduced here. It takes
into account only those entangled states that spread over the
whole system, and disregards any entanglement supported in a
smaller subsystem, very much like 3-tangle did. For example,
absence of entanglement in two component system
$\mathcal{H}_A\otimes\mathcal{H}_B$ for
$\dim\mathcal{H}_A\ne\dim\mathcal{H}_B$ reflects the fact that in
this case every state belongs to a smaller subspace $V_A\otimes
V_B$, $V_A\subset\mathcal{H}_A$, $V_B\subset\mathcal{H}_B$ as it
follows from Schmidt decomposition (\ref{Schmidt}). Entanglement
of such states should be treated in the corresponding  subsystems.
\end{rem}

\subsection{Hilbert-Mumford criterion}\label{HilMum}
The above examples, based on Theorem \ref{classic}, shows that
invariants are essential for understanding and quantifying of
entanglement. Unfortunately finding invariants is
%extremely difficult
a tough job, and more then 100 years of study give no hope for a
simple solution.

There are few cases where all invariants are known, some of them
were mentioned above.
%The only
%invariant of three-qubits is Cayley's hyperdeterminant
%$\mathrm{Det}(\psi)$ \cite{Miy}.
In addition invariants and covariants of four qubits and three
qutrits were found recently \cite{LT02, BLTh, BLThV}. For five
qubit only partial results are available  \cite{LT05}. See more on
invariants of qubits in \cite{OstSie04, OstSie05}.
%as well as classification of complex orbits \cite{VDM}.
%See also \cite{Perv} for invariants of system of format
For system of format $4\times4\times2$ the invariants are given in
\cite{Perv}.

Spin systems have an equivalent description in terms of {\it
binary forms\/}, see Example \ref{ent_spin}. Their invariants
%of
%spin states
%are known
are described by theory of {\it Binary Quantics\/}, diligently
pursued by mathematicians from the second half of 19-th century.
This is an amazingly difficult job, and complete success was
achieved by classics for $s\le 3$, the cases $s=5/2$ and $3$ being
one of the crowning glories of the theory \cite{Mum}. Modern
authors advanced it up to $s= 4$.

Other classical results of invariant theory are still waiting
physical interpretation and applications. In a broader context
{Bryce~S.~DeWitt} described the situation as follows:
\begin{quote}{\it ``Why should we not go directly to
invariants? The whole of physics is contained in them. The answer
is that it would be fine if we could do it. But it is not easy.''}
%\hfill {\color{blue} De Witt}
\end{quote}
Now, due to Hilbert's insight, we know that the difficulty
%stems from
is rooted in a perverse desire to put geometry into Procrustean
bed of algebra. He created {\it Geometric Invariant Theory}
%at the end of 19th century
just to overcome it.

\begin{thm}[Hilbert-Mumford  Criterion \cite{Mum}]
State $\psi\in\mathcal{H}$ is {\it entangled\/} iff every
observable $X\in\frak{L}={\rm Lie}(G)$ of the system in state
$\psi$ assumes
%both
a nonnegative
%and nonpositive
value with positive  probability.
\end{thm}
By changing $X$ to $-X$ one deduces that $X$ should assume
nonpositive values as well. So in entangled state no observable
can be biased  neither to strictly positive nor to strictly
negative values. Evidently completely entangled states with zero
expectations $\langle\psi|X|\psi\rangle=0$ of all observables pass
this test.
\begin{Ex}
Let $X=X_A\otimes1+1\otimes X_B$ be observable of two qubit system
%For two qubit system
$\mathcal{H}_A\otimes\mathcal{H}_{B}$ with
$${\rm Spec\,}X_A=\pm\alpha,\quad {\rm Spec\,}X_B=\pm\beta,\quad \alpha\ge\beta\ge0. $$
%Then
Suppose that $\psi$ is {\it unstable\/} and observable $X$ assumes
only strictly positive values in state $\psi$. Since those  values
are $\alpha\pm\beta$ then the state is decomposable
% In state $\psi$
%the observable $X$ assumes those values $\pm\alpha\pm\beta$ for
%which $\langle\psi|\pm\alpha\pm\beta\rangle\neq0$. Here
%$|\pm\alpha\pm\beta\rangle=|\pm\alpha\rangle\otimes|\pm\beta\rangle$
%%stands for
%denotes eigenstate of $X$ with eigenvalue $\pm\alpha\pm\beta$. If
%all assumed values  are strictly positive then the state $\psi$ is
%decomposable
$$\psi=a|\alpha\rangle\otimes|\beta\rangle+b|\alpha\rangle\otimes
|-\beta\rangle=|\alpha\rangle\otimes(a|\beta\rangle+b|-\beta\rangle),$$
% One can easily  check that
%if all the values are strictly positive then $\psi$ is separable,
i.e. Hilbert-Mumford criterion characterizes entangled qubits.
%the
%corresponding coefficient of decomposition $\psi=\sum_\pm
%a_{\pm\pm}|\pm\pm\rangle$ is nonzero.
\end{Ex}
The general form of H-M criterion may shed some light on the
nature of entanglement. However, it was
%intended
originally designed for application to {\it geometrical objects},
like linear subspaces or algebraic varieties of higher degree, and
its
%effectiveness
efficacy entirely depends on our ability to
%spell it out
express it in geometrical terms.
%of their geometric properties.
Let's give an example.
\begin{Ex}\label{ent_spin} {\it Stability of spin states.}
%Let $\mathcal{H}$ be Hilbert space of spin
%$1/2$ system, so that $\mathrm{SU\,}(\mathcal{H})=\mathrm{SU\,}(2)$ is spin group.
Spin $s$ representation $\mathcal{H}_s$
%of $\mathrm{SU\,}(2)$
%is isomorphic to
can be realized in space of {\it binary forms\/}  $f(x,y)$ of
degree $d=2s$
$$\mathcal{H}_s=\{f(x,y)|\deg f=2s\}$$
in which $\mathrm{SU\,}(2)$ acts by linear substitutions
$f(x,y)\mapsto f(ax+by,cx+dy)$. To make swap  from physics to
mathematics easier we denote by $f_\psi(x,y)$ the form
corresponding to state $\psi\in\mathcal{H}_s$. Spin state
$\psi\in\mathcal{H}_s$ can be treated algebraically, physically,
or geometrically according to the following equations
%binary form Here are three useful representations of binary form
%We write cast them in one the form
$$
\psi=\sum_{\mu=-s}^{\mu=s}a_\mu{2s\choose s+\mu}x^{s+\mu}y^{s-\mu}
=\sum_{\mu=-s}^{\mu=s}a_\mu{2s\choose s+\mu}^{1/2}|\mu\rangle\\
=\prod_i(\alpha_i x-\beta_i y).
$$
The first one is purely algebraic, the second
%expresses $\psi$ in terms of eigenstates
gives physical decomposition over eigenstates
$$|\mu\rangle={2s\choose s+\mu}^{1/2}x^{s+\mu}y^{s-\mu},\quad J_z|\mu\rangle=\mu|\mu\rangle$$
of spin projector operator
$J_z=\frac12\left(x\frac{\partial}{\partial
x}-y\frac{\partial}{\partial y}\right)$,
%with spin projection $\mu$,
and the last one is geometrical. It describes   form $f_\psi(x,y)$
in terms of configuration of its roots $z_i=(\beta_i:\alpha_i)$ in
{\it Riemann sphere\/} $\mathbb{C}\cup\infty=\mathbb{S}^2$ (known
also as {\it Bloch sphere\/} for spin $1/2$ states, and {\it
Poincar\'e sphere\/} for polarization of light).

According to H-M criterion state $\psi$ is {\it unstable\,} iff
spin projections onto some direction $\ell$ are strictly positive.
%its decomposition over eigenstates $|\mu\rangle$ of a spin
%projector operator $J_\ell$ contains only states with $\mu>0$.
By
%unitary
rotation we reduce the problem to $z$-component
$J_z=\frac12\left(x\frac{\partial}{\partial
x}-y\frac{\partial}{\partial y}\right)$ in which case the
corresponding form
$$f_\psi(x,y)=\sum_{\mu>0}a_\mu{2s\choose s+\mu}^{1/2}|\mu\rangle=\sum_{\mu>0}a_\mu{2s\choose
s+\mu}x^{s+\mu}y^{s-\mu}$$ has root $x=0$ of multiplicity more
then $s=d/2$. As result we arrive at the following criterion of
entanglement (=semistability) for spin states
\begin{equation}\label{mult}%\framebox{$
\psi\mbox{ is entangled}\Longleftrightarrow
\mbox{no more then half of the roots $f_\psi(x,y)$ coinside.}
%}$}
\end{equation}
One can show that if {\it less\,} then half of the roots coincide
then the state is {\it stable\/} i.e. can be transformed into a
completely entangled one by Lorentz group
$\mathrm{SL\,}(2,\mathbb{C})$ acting on roots
$z_i\in\mathbb{C}\cup\infty$ by M\"obius transformations
$z\mapsto\frac{az+b}{cz+d}$. In terms of these roots entanglement
equation (\ref{ent_eqn}) amounts to the following condition
\begin{equation}\label{balanced}%\framebox{$
\psi\mbox{ completely
entangled}\Longleftrightarrow
\sum_i(z_i)=0,
%$}
\end{equation}
where parentheses denote {\it unit vector\/}
$(z_i)\in\mathbb{S}^2\subset\mathbb{E}^3$ mapping into
$z_i\in\mathbb{C}\cup\infty$ under stereographic projection. For
example, for integral spin completely entangled state $|0\rangle$
%from balanced configuration (\ref{balanced}) consisting from $s$
can be obtained by putting equal number of points at the North and
the South poles of Riemann sphere. Another balanced configuration
(\ref{balanced}) consisting of $2s$ points evenly distributed
along the equator
%to get a balanced configuration (\ref{balanced})
produces completely entangled  state
$|\psi\rangle=\frac{1}{\sqrt{2}}(|s\rangle-|-s\rangle)$ for any
$s\ge 1$, cf. Example \ref{spin1_var}.

Note also that a configuration with half of its points in the
South pole can't be transformed into a balanced one
(\ref{balanced}), except all the remaining points are at the
North. However this can be done asymptotically by homothety
$z\mapsto
\lambda z$ as $\lambda
\rightarrow \infty$ which sends all points except zero to
infinity. This gives an example of  {\it semistable\/} but {\it
not\/} stable configuration.
%, corresponding to the North pole.
%points at the North pole, and for integral spin $s$ by taking
%balanced configuration
%mean
%stereographic projection of $z_i\in\mathbb{C}\cup\infty$ into unit
%vector $(z_i)\in\mathbb{S}^2\subset\mathbb{E}^3$.

%This is an illustration of general principle how
{\it Summary.}
%Thus
Solvability of the {\it nonlinear problem\/} of conformal
transformation of a given configuration into a balanced one
%configuration $z_i\in
%\mathbb{S}^2$ into a balanced one
(\ref{balanced})
%is  governed by
depends on  {\it topological condition\/} (\ref{mult}) on its
multiplicities. One can find application of this principle to
quantum marginal problem in \cite{Kl04, Kl05}.

%\marginpar
%{\footnotesize
%Actually solvability of a nonlinear equation usually governed by
%topological conditions, e.g. the number of zeros of a polynomial
%in a domain is governed by winding index (Rouche thm), Lefschetz
%trace formula for number of fixed points, zeros of vector fields
%etc.}

%We  apply a similar approach,
%%based on Geometric Invariant Theory,
%to a nonlinear {\it quantum marginal problem\/}
%%to be discussed
%in the next section.
%We'll give more advanced  examples of such topological conditions
%of solvability
%%based on geometric invariant theory
%in the next section. They all originate in Geometric Invariant
%Theory.
%In the next section we give more substantial examples of such
%More advanced examples of topological conditions for solvability
%of nonlinear problems will be given in the next section.
% of points $z_i\in
%\mathbb{S}^2$ governs  solvability of nonlinear problem of
%transforming this the configuration into a balanced one
%(\ref{balanced}) by a M\"obius transformation.
% Completely entangled spin
%states also can be described in term of the configuration roots in
%Riemann sphere.
%A commonly known
%invariant of the form is discriminant
%$$\Delta(f)=\prod_{i<j}(\alpha_i\beta_j-\alpha_j\beta_i),\qquad \Delta(f)\ne 0\Leftrightarrow p_i\ne p_j.$$
%Hence by Theorem \ref{classic} form $f(x,y)$ with no multiple
%roots is semistable (=entangled).
%$\mu\equiv s \mod 1$.
\end{Ex}

\bibliography{Q_Margins}
\bibliographystyle{plain}
\end{document}